\documentclass[twocolumn,traditabstract]{aa}
\usepackage[nonamebreak]{natbib}
\usepackage[stable]{footmisc}
\usepackage[fleqn]{amsmath} 
\usepackage{txfonts}
\usepackage{placeins}
\usepackage{natbib}
\bibpunct{(}{)}{;}{a}{}{,}
\usepackage{graphicx}
\usepackage{epstopdf}
\usepackage{ifthen}
\usepackage{mathtools}
\usepackage[breaklinks, colorlinks, citecolor=blue]{hyperref}
\usepackage{url}
\usepackage{comment}
\usepackage[table,usenames,dvipsnames]{xcolor}
\usepackage{xcolor}
\usepackage{soul}

\usepackage{datetime} 

\def\setsymbol#1#2{\expandafter\def\csname #1\endcsname{#2}}
\def\getsymbol#1{\csname #1\endcsname}

\def\Planck{\textit{Planck}}

\newbox\tablebox    \newdimen\tablewidth
\def\leaderfil{\leaders\hbox to 5pt{\hss.\hss}\hfil}
\def\tablenote#1 #2\par{\begingroup \parindent=0.8em
    \abovedisplayshortskip=0pt\belowdisplayshortskip=0pt
    \noindent
    $$\hss\vbox{\hsize\tablewidth \hangindent=\parindent \hangafter=1 \noindent
    \hbox to \parindent{$^#1$\hss}\strut#2\strut\par}\hss$$
    \endgroup}

\def\L2{\ifmmode L_2\else $L_2$\fi}

\def\DeltaT{\ifmmode \Delta T\else $\Delta T$\fi}
\def\deltat{\ifmmode \Delta t\else $\Delta t$\fi}
\def\fknee{\ifmmode f_{\rm knee}\else $f_{\rm knee}$\fi}
\def\Fmax{\ifmmode F_{\rm max}\else $F_{\rm max}$\fi}
\def\solar{\ifmmode{\rm M}_{\mathord\odot}\else${\rm M}_{\mathord\odot}$\fi}
\def\Msolar{\ifmmode{\rm M}_{\mathord\odot}\else${\rm M}_{\mathord\odot}$\fi}
\def\Lsolar{\ifmmode{\rm L}_{\mathord\odot}\else${\rm L}_{\mathord\odot}$\fi}
\def\inv{\ifmmode^{-1}\else$^{-1}$\fi}
\def\mo{\ifmmode^{-1}\else$^{-1}$\fi}
\def\sup#1{\ifmmode ^{\rm #1}\else $^{\rm #1}$\fi}
\def\expo#1{\ifmmode \times 10^{#1}\else $\times 10^{#1}$\fi}
\def\,{\thinspace}
\def\lsim{\mathrel{\raise .4ex\hbox{\rlap{$<$}\lower 1.2ex\hbox{$\sim$}}}}
\def\gsim{\mathrel{\raise .4ex\hbox{\rlap{$>$}\lower 1.2ex\hbox{$\sim$}}}}

\def\simprop{\mathrel{\raise .4ex\hbox{\rlap{$\propto$}\lower 1.2ex\hbox{$\sim$}}}}
\def\deg{\ifmmode^\circ\else$^\circ$\fi}
\def\pdeg{\ifmmode $\setbox0=\hbox{$^{\circ}$}\rlap{\hskip.11\wd0 .}$^{\circ}
          \else \setbox0=\hbox{$^{\circ}$}\rlap{\hskip.11\wd0 .}$^{\circ}$\fi}
\def\arcs{\ifmmode {^{\scriptstyle\prime\prime}}
          \else $^{\scriptstyle\prime\prime}$\fi}
\def\arcm{\ifmmode {^{\scriptstyle\prime}}
          \else $^{\scriptstyle\prime}$\fi}
\newdimen\sa  \newdimen\sb
\def\parcs{\sa=.07em \sb=.03em
     \ifmmode \hbox{\rlap{.}}^{\scriptstyle\prime\kern -\sb\prime}\hbox{\kern -\sa}
     \else \rlap{.}$^{\scriptstyle\prime\kern -\sb\prime}$\kern -\sa\fi}
\def\parcm{\sa=.08em \sb=.03em
     \ifmmode \hbox{\rlap{.}\kern\sa}^{\scriptstyle\prime}\hbox{\kern-\sb}
     \else \rlap{.}\kern\sa$^{\scriptstyle\prime}$\kern-\sb\fi}
\def\ra[#1 #2 #3.#4]{#1\sup{h}#2\sup{m}#3\sup{s}\llap.#4}
\def\dec[#1 #2 #3.#4]{#1\deg#2\arcm#3\arcs\llap.#4}
\def\deco[#1 #2 #3]{#1\deg#2\arcm#3\arcs}
\def\rra[#1 #2]{#1\sup{h}#2\sup{m}}

\def\dots{\relax\ifmmode \ldots\else $\ldots$\fi}
\def\WHzsr{\ifmmode $W\,Hz\mo\,sr\mo$\else W\,Hz\mo\,sr\mo\fi}
\def\mHz{\ifmmode $\,mHz$\else \,mHz\fi}
\def\GHz{\ifmmode $\,GHz$\else \,GHz\fi}
\def\mKs{\ifmmode $\,mK\,s$^{1/2}\else \,mK\,s$^{1/2}$\fi}
\def\muKs{\ifmmode \,\mu$K\,s$^{1/2}\else \,$\mu$K\,s$^{1/2}$\fi}
\def\muKRJs{\ifmmode \,\mu$K$_{\rm RJ}$\,s$^{1/2}\else \,$\mu$K$_{\rm RJ}$\,s$^{1/2}$\fi}
\def\muKHz{\ifmmode \,\mu$K\,Hz$^{-1/2}\else \,$\mu$K\,Hz$^{-1/2}$\fi}
\def\MJysr{\ifmmode \,$MJy\,sr\mo$\else \,MJy\,sr\mo\fi}
\def\MJysrmK{\ifmmode \,$MJy\,sr\mo$\,mK$_{\rm CMB}\mo\else \,MJy\,sr\mo\,mK$_{\rm CMB}\mo$\fi}
\def\microns{\ifmmode \,\mu$m$\else \,$\mu$m\fi}

\def\muK{\ifmmode \,\mu$K$\else \,$\mu$\hbox{K}\fi}
\def\microK{\ifmmode \,\mu$K$\else \,$\mu$\hbox{K}\fi}
\def\muW{\ifmmode \,\mu$W$\else \,$\mu$\hbox{W}\fi}
\def\kms{\ifmmode $\,km\,s$^{-1}\else \,km\,s$^{-1}$\fi}
\def\kmsMpc{\ifmmode $\,\kms\,Mpc\mo$\else \,\kms\,Mpc\mo\fi}
\providecommand{\sorthelp}[1]{}

\def\aap{{A\&A}}

\def\apjs{{ApJSupp.}}

\newcommand{\planck}{\Planck}

\newcommand{\srolltwo}{{\tt SRoll2} }
\newcommand{\focus}{{\tt FoCUS} }
\newcommand{\cwst}{CWST}
\newcommand{\cswst}{CWST }

\usepackage{color}
\usepackage{xcolor}

\def\be{\begin{equation}}
\def\ee{\end{equation}}
\def\ba{\begin{eqnarray}}
\def\ea{\end{eqnarray}}

\def\Planck{\textit{Planck}}

\begin{document}

\newcommand{\jmcomment}[1]{{\bf \color{green} (#1)}}
\newcommand{\fbcomment}[1]{{\bf \color{purple} (#1)}}
\newcommand{\EA}[1]{{\bf \color{blue} (#1)}}
\newcommand{\TREA}[1]{{\bf \color{red} (TREA: #1)}}
\newcommand{\egcomment}[1]{{\bf \color{orange} (#1)}}

\title{\vglue -10mm  Non-Gaussian modelling and statistical denoising of Planck  dust polarization full-sky maps  using scattering transforms \\}

\author{\small
J.-M.~Delouis\inst{1}~\thanks{Corresponding author: J.-M.~Delouis, jean.marc.delouis@ifremer.fr}
\and
E.~Allys\inst{2}
\and 
E.~Gauvrit\inst{1,2}
\and
F.~Boulanger\inst{2}
}

\institute{\small
Laboratoire d'Oc{\'e}anographie Physique et Spatiale (LOPS), Univ. Brest, CNRS, Ifremer, IRD, Brest, France\goodbreak
\and
Laboratoire de Physique de l’$\acute{\rm E}$cole Normale Sup$\acute{\rm e}$rieure, ENS, Universit$\acute{\rm e}$ PSL, CNRS, Sorbonne Universit$\acute{\rm e}$, Universit$\acute{\rm e}$ Paris Cit$\acute{\rm e}$, 75005 Paris, France
}

\date{\vglue -1.5mm \today \vglue -5mm}

\abstract{\vglue -3mm 
Scattering transforms have been successfully used to describe dust polarization for flat-sky images.  
This paper expands this framework to noisy observations on the sphere with the aim of obtaining denoised Stokes $Q$ and $U$ all-sky maps at 353\,GHz, as well as a non-Gaussian model of dust polarization, from the \Planck\ data. To achieve this goal, we extend the computation of scattering coefficients to the Healpix pixelation and introduce cross-statistics that allow us to make use of half-mission maps as well as the correlation between dust temperature and polarization. Introducing a general framework, we develop an algorithm that uses the scattering statistics to separate dust polarization from data noise. The separation is validated on mock data, before being applied to the \srolltwo \Planck\ maps at $N_\text{side} = 256$. The validation shows that the statistics of the dust emission, including its non-Gaussian properties, are recovered until $\ell_{\rm max} \sim 700$, where, at high Galactic latitudes, the dust power is smaller than that of the dust by two orders of magnitude. 
On scales  where the dust power is lower than one tenth of that of the noise, structures in the output maps have comparable statistics but are not spatially coincident with those of the input maps. 
Our results on \Planck\ data are significant milestones opening new perspectives for statistical studies of dust polarization and for the simulation of Galactic polarized foregrounds. The \Planck\ denoised maps is available (see \href{http://sroll20.ias.u-psud.fr/sroll40_353_data.html}{here.}) together with results from our validation on mock data, which may be used to quantify uncertainties.
 } 
 

\authorrunning{...}

\titlerunning{Model Planck polarized dust maps using Poor Wavelet Scattering Transform.}

\maketitle
\section{Introduction}
\label{sec:intro}

The cosmic microwave background (CMB) is a prime observational probe for constraining cosmological models~\citep{Durrer15}.
Today, with the uncertainties in the CMB temperature spectrum essentially reduced to the cosmic variance~\citep{planck2018_V,Planck2018XI}, the CMB experiments have shifted their focus to polarization.
In particular, accurate measurements of the tensor component ($B$-modes) of the polarized signal could provide direct evidence of the inflation period~\citep{Guth1981, LINDE1982}. This paramount goal of cosmology is 
driving the development of ambitious CMB experiments~\citep{cmbs4,Simons19,litebird2022}, but the potential detection of primordial B-modes does not only depend on increasing the signal-to-noise ratio on CMB polarization. 

The quest for CMB $B$-modes is also hampered by instrumental systematic effects~\citep{Planck2018_III} and polarized foregrounds dominated by Galactic dust emission~\citep{pb2015,Planck2018XI}. In this context, the modelling of systematic effects and Galactic foregrounds must advance alongside the sensitivity of the measurements. This is a major challenge because instrumental systematics and Galactic emission are non-Gaussian signals, which in essence are difficult to model. 
To address this difficulty, the CMB community has been investing much effort in the development 
of Galactic emission models~\citep{2013psm,Thorne17,Vans2017,Martinez18,Zonca21,Hervias22} combined with instruments models to produce end-to-end simulations of the data, as 
done e.g. for \Planck\ \citep{Planck2018_III}. 
Data simulations are essential to marginalise the inference of cosmological parameters on 
the nuisance signals and correct for bias~\citep[e.g.,][]{Vacher22}, and to perform Likelihood-Free Inference methods~\citep{PlanckLowEll,Alsing19,Jeffrey22}. 

Non-Gaussianity is an important characteristic of Galactic foregrounds. 
To account for it, several authors have introduced machine learning algorithms~\citep{Aylor20, Krachmalnicoff20, Petroff20, Thorne21}, but these methods need to be trained. Thereby their use is hindered by the difficulty of building relevant training sets. Magnetohydrodynamics (MHD) simulations of the interstellar medium~\citep{Kritsuk18,Kim19,Pelgrims22} are useful to develop the methodology but they are far from reproducing the statistics of dust polarization with the accuracy required for CMB components separation. 

Another approach to model Galactic foregrounds is to rely on Scattering Transform statistics. These statistics were introduced in data science to discriminate non-Gaussian textures~\citep{mallat2012,bruna2013,cheng2021quantify}, and they have since be applied to dust emission maps computed from MHD simulations~\citep{Allys_2019, blancard2020statistical, Saydjari20}. Promising results have also been obtained on various astrophysical processes, as large scale structures density field and galaxy surveys~\citep{allys2020,eickenberg2022wavelet,valogiannis2022going,valogiannis2022towards}, weak-lensing convergence maps~\citep{cheng20,cheng2021weak}, and 21cm data of the epoch of reionization~\citep{greig2022exploring}. To construct these statistics, convolutions of the input image with wavelets over multiple oriented scales are combined with 
non-linear operators that allow to efficiently characterize interactions between scales.  

One notable advantage of the scattering transforms is that generative models reproducing quantitatively the non-Gaussian structures of a given process can be constructed from a small number of realizations of this process, which can be even a single image~\citep{bruna2019multiscale,allys2020}. This could allow to construct Galactic dust models directly from observational data. For this purpose, the \Planck\ data are a key observational input. They have been used both as a template of the dust sky and to model the spectral energy distribution of dust emission~\citep{Thorne17,Zonca21}. However, for polarization, data noise is a severe limitation that must be circumvented. A new direction was opened by~\citep{Regaldo2021} who introduced an algorithm successfully using scattering statistics to separate dust emission from data noise. They applied it to flat-sky \Planck\ Stokes images at 353\,GHz. Their method exploits the very different non-Gaussian structure on the sky of dust emission compared to the data noise. An iterative optimization on sky pixels yields denoised maps and a generative model of dust polarization. 

This paper aims to extend the innovative components separation approach of~\citep{Regaldo2021} to the sphere and to apply it to all-sky \Planck\ polarization maps. Our scientific motivation is to obtain denoised \Planck\ Stokes maps that may be used for the modelling of the dust foreground to CMB polarization. For the astrophysics of dust polarization, the signal-to-noise ratio limits the range of angular scales accessible to study~\citep{Planck2018XII}. Thus, our work is also a valuable contribution to statistical studies of dust polarization at high Galactic latitudes. 

Note that while our scientific objective is the statistical denoising of dust polarized emission, we treat this problem as the separation between two components. This differs from usual denoising algorithms that rely for instance on filtering~\citep{wiener1949extrapolation,zaroubi1994wiener} or sparsity~\citep{starck2002curvelet}
In comparison to these usual algorithms, our objective is to recover a map with the correct statistics, even if it differs from the true map at the smallest scales. We also emphasize that the methodology we present in this paper can be similarly applied to the separation of two components of interest, and not only for denoising.

Our data objective involves several challenges. The scattering coefficients must be computed on all-sky maps in Healpix format. The algorithm used to separate dust and data noise must accommodate variations in the dust statistics over the sky. The computing time required to compute the scattering coefficients and the map optimization must be kept manageable. Within this framework, we developed cross-scattering coefficients to make optimal use of the available data, especially complementary half-mission maps. cross-scattering coefficients have also been introduced by~\citep{Regaldo22} to model multi-channel dust data. In doing this, the two papers extend the commonly use of cross-power spectra to statistics that encode non-Gaussianity. For our project, the cross-correlation of \Planck\ maps with independent noise realizations facilitates the denoising. It also allows us to take into account the $TE$ and $TB$ correlation of dust polarization~\citep{Planck2018XI}.  

The paper is organized as follows. 
Section~\ref{sec:sec_method} introduces the cross-scattering statistics, the algorithm and the loss functions from a generic starting-point. Our method is validated on mock data (Sect.~\ref{sec:test_simu}) before being applied to the \Planck\ maps (Sect.~\ref{sec:result_data}). 
Applications and perspectives for future work are discussed in Sec.~\ref{sec:prospects}. The paper results are summarized in Sect.~\ref{sec:conclusion}. Additional figures are presented in Appendix~\ref{sec:AppendixA}.  
%

\section{CWST and dust/noise components separation}
\label{sec:sec_method}

We introduce our method in Sect.~\ref{sec:method}, before presenting the cross-scattering transform on the sphere in Sect.~\ref{sec:valcswst}. Next, we describe the algorithm we use to perform the components separation between dust polarization and data noise  (Sect.~\ref{sec:focus}).

\subsection{Introduction}
\label{sec:method}

The present study aims to characterize the statistical properties of the polarized dust emission from noisy \Planck\ all-sky maps in Healpix~\citep{healpix2005}.
We ignore the CMB, which can be either removed or neglected, and address this problem as a separation between two components: dust emission and data noise.  
We choose to work with \srolltwo \Planck\ polarization maps at 353\,GHz~\citep{delouis2019}. We convert the all-sky Stokes $Q$ and $U$ maps 
into $E$ and $B$ maps, and apply our method to the latter because they are independent scalars that do not depend on the chosen reference frame. This transformation being non-local, signal from the brightest areas in the Galactic plane contaminate the $E$ and $B$-maps at high Galactic latitude. Thus, to display the results we transform the denoised $E$ and $B-$maps back to $Q$ and $U$. In doing so, we also conform to the standard way of representing the polarised sky. In practice, we experimented the use of $E$ and $B$ is better for the present work. 

Our data processing is the same for $E$ and $B$ maps. In each case, we have access to a full-mission map $d$, and two half-missions $d_1$ and $d_2$.The half-missions maps are computed from respectively the first and second part of the mission, making their noises and time variable instrumental systematics mostly independent.
These maps can be written as a sum of the dust emission $s$ and three different noise realisations, that we call respectively $n$, $n_1$ and $n_2$. One has for instance
\begin{eqnarray}
\label{eq:datamodel}
d &=& s+n,
\end{eqnarray}
for the full mission, and
\begin{eqnarray}
\label{eq:datamodel2}
d_1 &=& s+n_1,
\end{eqnarray}
for the first half mission. We assume that the noises for the two half-missions can be considered statistically independent, at least for $\ell>30$ (see for instance~\cite{delouis2019}). Thanks to a simulation effort, we also assume that we have access to a large number of realistic noise maps. The \srolltwo dataset contains for instance an ensemble of 500 $(\tilde{n},\tilde{n}_1,\tilde{n}_2)$ associated triplets of maps~\citep{delouis2019}, which can be used to simulate the noises of the full-mission and of both half-missions of $E$, $B$, and $T$ maps. Finally, we also have access to an intensity map, that we label $T$, and that will be used to characterize the statistical dependency between polarized and total intensity dust emission. The details of the different data used and the associated assumptions will be discussed in Sec.~\ref{sec:test_simu} and~\ref{sec:result_data}.

To obtain the statistical properties of the polarized dust emission, we will perform a components separation of the dust and noise maps, using their different non-Gaussian characterization as a lever arm~\citep{blancard2020statistical}. This components separation generates a new dust maps $\tilde{s}$ from a gradient descent with an ensemble of statistical constraints built from Cross Wavelet Scattering Transforms (CWST) between different maps. The statistics of the polarized dust map $s$ is then estimated from this $\tilde{s}$ map. In the following, we call \focus this components separation method, as Function Of Cleaning Using Statistics. 

\subsection{Cross Wavelet Scattering Transform}
\label{sec:valcswst}

Scattering Transforms (ST) are a recently developed type of non-Gaussian summary statistics~\citep{mallat2012,bruna2013}. They are inspired from convolutional neural networks, but do not need any training stage to be computed (\textit{i.e.}, the function to compute the statistics from a set of data can be written explicitly, and do not need to be learnt). They thus benefit for the low-variance efficient characterization typical to neural networks, but give some level of interpretability through their explicit mathematical form. Several sets of ST statistics have been constructed, as the Wavelet Scattering Transform (WST~\cite{Allys_2019,blancard2020statistical}, or the Wavelet Phase Harmonics (WPH~\cite{phase2018,allys2020}). The construction of the ST statistics relies on two main features: scales separation (through wavelet transforms), and characterization of the interaction between scales using non-linearity (as modulus, ReLU, or phase acceleration). The Cross Wavelet Scattering Transform (CWST), introduced in this paper, is a new type of ST cross-statistics constructed for data defined on the sphere. It is an extension of the WST, to which it boils down when use as auto-statistics.

\begin{figure}[!ht]
\centering
\includegraphics[width=0.5\textwidth]{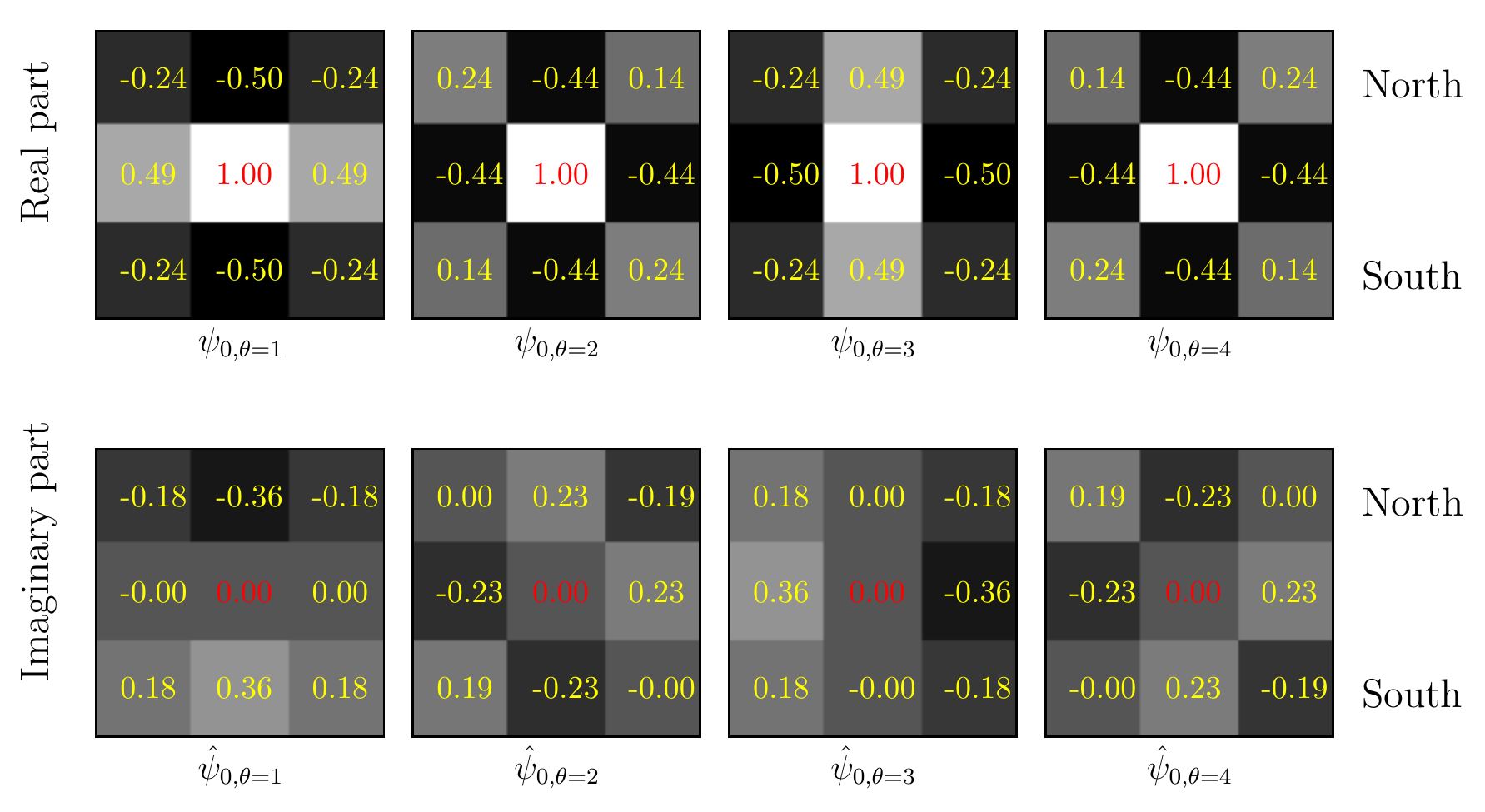}
\caption{Representation of the four $\tilde{\psi}_{0,\theta}$ 3x3 kernels. The first row shows the real part of the kernel coefficients while the second row shows there imaginary parts. The right labels show how the coefficients are applied to the Healpix neighbours. In this figure, top corresponds to North (as defined by Healpix), left to West, and so on.}
\label{fig:pwst_coef}
\end{figure}

The first building block of the CWST transform is a wavelet transform allowing fast computation directly on HealPix data. For this purpose, we introduced a very simple multi-resolution wavelet transform defined from four 3x3 complex kernels, which are used to compute convolutions with HealPix maps at different resolutions. These kernels, which are called $\tilde{\psi}_{j,\theta}$ with $\theta$ between 1 to 4, are plotted in Fig.~\ref{fig:pwst_coef}.With respect to the Healpix conventions, $\theta=1$ refers to a North-South brightness oscillation associated to an East-West elongated structure. The convolution is computed by multiplying the weights of the Fig.~\ref{fig:pwst_coef} to the 8 neighbours of each pixel in such way that top left value is related to the North-West Healpix pixel\footnote{In the particular case where a pixel has only 7 neighbours, the value taken for the 8th missing pixel corresponds to the one of the closest anti-clockwise pixel (e.g. North-West to replace North). With a resolution $N_{side}$ it occurs one pixel every ${N_{side}}^2$ pixels.}.

The different wavelets $\psi_{j,\theta}$ are labeled by a integer scale $j$ going from 0 to $J-1$, and by a integer angle $\theta$ going from 1 to $L$ (associated to a $\left(\theta-1\right) \cdot \pi/L$), for a total of $J\cdot L$ wavelets. The present work on $N_\text{side} = 256$ map uses $J=8$ and $L=4$ (since the four $\tilde{\psi}_{j,\theta}$ kernels define 4 orientations). The wavelet transform for $j=0$ is obtained through the convolution of the input map in Healpix with the $\tilde{\psi}_{0,\theta}$ kernels. The wavelet transform for $j=1$ is then obtained by first sub-sampling the input map by computing 2x2 mean, thanks to the Healpix nested indexing property, and computing again the convolutions with the $\tilde{\psi}_{0,\theta}$ kernels. By repeating this process, one can compute convolutions up to scales $j=J-1$.

As the $\tilde{\psi}_{0,\theta}$ kernels have a 2 pixels wavelength, the characteristic pic multipole probed by the wavelet transform at scale $j$ is $\ell \approx 512 \cdot 2^{-j}$. Indeed, starting from an initial resolution of $N_\text{side} = 256$ at $j=0$, the resolution on which the convolution is done at $j > 0$ is $N_\text{side} = 256 \cdot 2^{-j}$. One also sees that the initial value of $N_\text{side}$ obviously limits the number of scales which can be considered: here $j=7$, which correspond to an Healpix map with $N_{side}=2$. 

We are aware of the fact that more refined way to compute wavelet transform on the sphere exist~\citep[see e.g.,][]{leistedt2013s2let,mcewen2015directional,mcewen2018localisation}, from which a first implementation of scattering transform on the sphere has already been defined~\citep{mcewen2021scattering}. We have chosen to do this project within our scheme because of its simplicity of use, as well as its computational efficiency for GPU accelerated computations, especially for the implementation of new cross-statistics. We however would like to transition to better wavelets in the future, for which we could expect an improvement of the obtained results.

In the following, we use an index $\lambda = (j,\theta)$ to describe the oriented scale associated to each $\psi_\lambda$, and we keep implicit the fact that they are defined on HealPix maps of different $N_\text{side}$. The wavelet convolution of an image $I$ with a $\psi_{\lambda}$ wavelet then reads $I \star \psi_{\lambda}(p)$, where $p$ is the coordinate on the sphere at the corresponding resolution.

The CWST cross-statistics are calculated on two maps $I_a$ and $I_b$. Similarly to the usual WST, the CWST contain two layers of coefficients, which are characterized respectively by one or two oriented scales $\lambda_i$. The whole set of statistics, called $S(I_a,I_b)$, is thus decomposed in the $S_1$ coefficients at first layer, and the $S_2$ coefficients at second layer. Note that when $I_a = I_b$, these coefficients boil down to the usual WST coefficients~\citep{bruna2013}. 

The coefficients at first order are called $S_1(I_a,I_b)_{\lambda_1}$. They are constructed from a product of convolutions of $I_a$ and $I_b$ at the same $\lambda_1$ scale:
\begin{multline}
\label{eq:c1pwst}
{C_1}( I_a, I_b)^\Re_{\lambda_1}= \text{sign}\left(\Re\left(I_{a}\star \psi_{\lambda_1}\cdot {I_{b}\star \psi_{\lambda_1}} ^*\right)\right) \\
\cdot \sqrt{\left\lvert \Re\left(I_{a}\star \psi_{\lambda_1}\cdot {I_{b}\star \psi_{\lambda_1}} ^*\right)\right\rvert },
\end{multline}
and
\begin{multline}
{C_1}( I_a, I_b)^\Im_{\lambda_1} = \text{sign}\left(\Im\left(I_{a}\star \psi_{\lambda_1}\cdot {I_{b}\star \psi_{\lambda_1}} ^*\right)\right) \\
\cdot \sqrt{\left\lvert \Im\left(I_{a}\star \psi_{\lambda_1}\cdot {I_{b}\star \psi_{\lambda_1}} ^*\right)\right\rvert }
\end{multline}
where we considered independently the real and imaginary part of the products of wavelet convolutions, and where $*$ and $|\cdot|$ stand respectively for the complex conjugate (acting here on the whole ${I_{b}}\star \psi_{\lambda_1}$ term) and absolute value. The square root allows to recover L1-like norm, which is useful to decrease the variance of the estimators, but introduces a bias when computing the correlated information between two noisy data sets. 

The $S_1$ coefficients are then computed from a spatial integration:
\begin{eqnarray}
\label{eq:s1pwst}
S_1(I_a,I_b)_{\lambda_1} &=& \left\langle C_1( I_a, I_b)^\Re_{\lambda_1} + i \cdot C_1( I_a, I_b)^\Im_{\lambda_1} \right\rangle_\text{pixels},
\end{eqnarray}
where $i$ is the imaginary unit, and where the brackets stand for a spatial average on the sphere, which is multiplied by $2^{j}$ to have a uniform normalisation\footnote{Meaning that the $S_1$ coefficients of a white noise are constant across scales.}. The $S_1$ coefficients can also be used to compute the statistics of a single map $I_a=I_b=I$. In this case, the $C^\Im_{\lambda_1}$ term obviously vanishes, the $C^\Re_{\lambda_1}$ terms boils down to the complex modulus of the $I\star \psi_{\lambda_1}$ convolution, and one recovers for $S_1$ the standard WST definition $S_1=\langle | I\star \psi_{\lambda_1}| \rangle_\text{pixel}$. 

The coefficients at second order are called $S_2(I_a,I_b)_{\lambda_1,\lambda_2}$. They are constructed from two positive and negative terms:
\begin{eqnarray}
\label{eq:s2\cwst_bis}
C_{2,+}^{\Re}(I_{a}, I_{b})_{\lambda_1,\lambda_2} &=& \left|\psi_{\lambda_2} \star ReLU \left(C_1^\Re( I_{a}, I_{b})_{\lambda_1} \right)\right|, \\ \nonumber
C_{2,-}^{\Re}(I_{a}, I_{b})_{\lambda_1,\lambda_2} &=& \left|\psi_{\lambda_2} \star ReLU \left(-C_1^\Re( I_{a}, I_{b})_{\lambda_1} \right)\right| ,
\end{eqnarray}
and similarly for the imaginary terms $C_{2,\pm}^\Im$. The $S_2$ terms are then obtained through a spatial integration:
\begin{multline}
\label{eq:s2\cwst}
{S}_{2}( I_{a}, I_{b})_{\lambda_1,\lambda_2} = \bigg\langle \left( C_{2,+}^{\Re}(I_{a}, I_{b})_{\lambda_1,\lambda_2} - C_{2,-}^{\Re}(I_{a}, I_{b})_{\lambda_1,\lambda_2} \right) \\
+ i \cdot \left( C_{2,+}^{\Im}(I_{a}, I_{b})_{\lambda_1,\lambda_2} - C_{2,-}^{\Im}(I_{a}, I_{b})_{\lambda_1,\lambda_2} \right) \bigg\rangle_\text{pixels}.
\end{multline}
For these coefficients, one also recover the standard WST definition $S_2=\langle || I \star \psi_{\lambda_1}| \star \psi_{\lambda_2}| \rangle_\text{pixel}$ when $I_a=I_b=I$.

To use an algebraic sum in Eq.~\eqref{eq:s2\cwst} allows to easily identify statistical dependencies between processes. Indeed, if the two images are correlated or anti-correlated, we expect the $C_1(I_{a}, I_{b})$ to be respectively mostly positive or negative, leading to $S_2$ values of the same sign. On the other hand, if the two images are uncorrelated, we expect $C_1(I_{a}, I_{b})$ to have similar positive and negative patterns, thus leading to $S_2$ coefficients of much lower values. In addition, using a convolution with a second wavelet allows to quantify at which scales such a correlation appears, and thus to characterize an interaction between two oriented scales.

\subsection{Principle of the components separation.}
\label{sec:focus}

We present here the components separation method between dust polarization and the data noise, that we call \focus for Function Of Cleaning Using Statistics. This method consists in generating a new dust map through a gradient-descent in pixel space under several constraints constructed from CWST statistics. In this part, we call $u$ the dust map which is modified in the gradient descent. This map is initialized by $d$, and we call $\tilde{s}$ its final value, which is the \focus dust map. The scientific result of this paper is this dust map, as well as its statistics. As will be discussed in the validation performed in Sec.~\ref{sec:test_simu}, while the statistics of $\tilde{s}$ reproduce very well the ones of the unknown $s$ dust map, its deterministic structures are not reproduced at the smallest scale.

The three constraints imposed to the $u$ map are constructed from the full and half-mission maps $d$, $d_1$, $d_2$, a $T$ temperature maps, as well as an ensemble of realizations of full and half-mission noises $\{\tilde{n}, \tilde{n}_1, \tilde{n}_2\}$. These constraints, which are not independent, are obtained from averages over the ensemble of noise realizations.

The first constraint is 
\begin{equation}
S(d_1,d_2) \simeq \left\langle S(u+\tilde{n}_1,u+\tilde{n}_2) \right\rangle_{\tilde{n}},
\end{equation}
where the bracket designs an average over the noise realizations, here $\{\tilde{n}_1,\tilde{n}_2\}$. This constraint enforces the statistics of $u$ to the statistics of $s$ estimated from the two half-mission maps.
The second constraint, which yields,
\begin{equation}
\label{EqConstraintSdu}
S(d,u) \simeq \left\langle S(u+\tilde{n},u) \right\rangle_{\tilde{n}},
\end{equation}
enforces the cross-statistics between $u$ and $d$.
Note the difference between both constraints: the first one is only statistical in nature, since it contains no cross-term between the denoised $u$ map and observational data, while the second one includes a cross-term between $u$ and $d$. This allows us to use cross-statistics between half-mission data that have independent noises for the first constraint, while we use the full-mission map, which has the smallest amount of noise, for the second one. This second term also constraints deterministic features. Indeed, similarly to a cross-spectrum computation, $S(d,u)$ characterizes the correlation between $d$ and $u$ (here a non-linear correlation), and hence the correlation between $s$ and $u$ if $n$ and $u$ are independent
This term thus constraints the alignment of structures\footnote{The $C_1$ terms for instance impose that local levels of oscillation at a given scale are correlated.} in $u$ with structures in $s$, which we call deterministic features.

Finally, the last loss constraint,
\begin{equation}
S(T,d) \simeq \left\langle S(T,u+\tilde{n}) \right\rangle_{\tilde{n}},
\end{equation}
imposes to keep the same cross-statistics between $T$ and $u$ than those estimated from the $d$ map. For this last constraint, we choose $T$ to be the 857\,GHz SRoll3 map, which is corrected for large scale systematics present in earlier data releases~\citep{Lopez2021}. The choice of $857\,\rm{GHz}$ instead of the $353\,\rm{GHz}$ maps avoid correlated noise between the polarization maps $d$ and the $T$ map (as they are computed using data from the same detectors for intensity and polarization). A drawback is that the $857\,\rm{GHz}$ $T$ map includes a significant contribution from the Cosmological Infrared Background at high Galactic latitudes, but this emission is only very weekly polarized \citep{Feng20,Lagache20}. 
Also, the non-negligible spatial variations of the spectral energy distribution of the Galactic dust induce some decorrelation between the $T$ maps at $857$ and $353\,\rm{GHz}$~\citep{dipole2021}. An advantage of our method is that it is not hindered by such effects, since we only impose the denoised map to have the same statistical dependency than the one estimated from the observational data.

Note that we also tried to impose constraints involving directly the recovered noise map, $d-u$. We considered in particular a direct constraint on its statistics, as well as on its independence with $u$. Both these constraints did not improve particularly the results obtained. We believe in particular that the independence between $u$ and $d-u$ is already constrained from Eq.~\eqref{EqConstraintSdu}. It is however possible that such constraints would play a non-negligible role to separate other types of astrophysical fields.

In practice, to perform a gradient descent from the above constraints would however be computationally very costly, due to the necessity to compute an average on a large number of noise realisations (300 \footnote{500 noise realizations were available but only 300 has been used to limit the memory and computing usage. We also checked that adding additional noises seems not to improve the results anymore.}) at each iteration. To avoid it, the noise-induced biases of the CWST statistics are separated on a specific term and estimated only after a certain batch of iteration. The loss term related to the first constraint is thus written
\begin{equation}
\label{eq:loss1}
\text{Loss}_{1} = {\left|\left|\frac{S(d_1,d_2) - S(u,u) - \text{B}_1 }{\sigma_{S(u+\tilde{n}_1,u+\tilde{n}_2)}}\right|\right|}^{2},
\end{equation} 
with
\begin{equation}
\label{eq:bias1}
\text{B}_1 = \big\langle S(u+\tilde{n}_1,u+\tilde{n}_2) - S(u,u) \big\rangle_{\tilde{n}},
\end{equation} 
where $||.||^2$ stands for the square Euclidean norm over the whole set of CWST statistics, and $\sigma$ stands for the estimated standard deviation of $\left\langle S(u+\tilde{n}_1,u+\tilde{n}_2) \right\rangle_{\tilde{n}}$. As discussed above, $B_1$ gives an estimated of the noise-induced bias between $S(d_1,d_2)$ and $S(s,s)$, which is more accurate the closer $u$ gets to $s$ during the optimization.

Similarly, the loss terms related to the second constraint yields:\footnote{In practice, the numerical experiments we did showed that it was difficult to use a loss term involving a difference between two $u$ terms, especially in the start of the optimization where $u_0 = d$. To avoid this, we modified this loss term, replacing $S(u,u)$ by $S(d_1,d_2) -B_1$, which proved much more efficient. This led to the following loss term:
\begin{equation}
\label{eq:loss2bis}
\text{Loss}_{2} = {\left|\left|\frac{S(d,u) - S(d_1,d_2) - \text{B}_2 + \text{B}_1}{\sqrt{\sigma_{S(u+\tilde{n},u)}^2+\sigma_{S(u+\tilde{n}_1,u+\tilde{n}_2)}^2}}\right|\right|}^{2}.
\end{equation}}
\begin{equation}
\label{eq:loss2}
\text{Loss}_{2} = {\left|\left|\frac{S(d,u) - S(u,u) - \text{B}_2}{\sigma_{S(u+\tilde{n},u)}}\right|\right|}^{2},
\end{equation} 
with
\begin{equation}
\label{eq:bias2}
\text{B}_2 = \big\langle S(u+\tilde{n},u) - S(u,u) \big\rangle_{\tilde{n}},
\end{equation} 
and the one related to the third constraint yields:
\begin{equation}
\label{eq:loss3}
\text{Loss}_{3} = {\left|\left|\frac{S(T,d) - S(T,u) - \text{B}_3 }{\sigma_{ S(T,u+\tilde{n}) }}\right|\right|}^{2},
\end{equation} 
with
\begin{equation}
\label{eq:bias3}
\text{B}_3 = \big\langle S(T,u+\tilde{n}) - S(T,u) \big\rangle_{\tilde{n}}.
\end{equation} 

These losses treat the Galactic signal as an homogeneous process on the sky. However, it is clear that the dust emission has a strong variation in statistical properties with Galactic latitude. To take this into account, the three losses described previously are computed from statistics estimated on different parts of sky, using 5 different standard Planck masks with $f_\text{sky} \in \left[1.0,0.73,0.63,0.43,0.27\right]$~\citep{PlanckXXX2016}. Dust statistics are dominated by the brightest emission within the unmasked sky. Therefore, as the amplitude of the dust emission decreases steadily from the Galactic plane to the poles, the masks allow us to progressively characterize dust polarization from bright to faint regions when $f_\text{sky}$ goes from $1.0$ to $0.27$. Since in contrast, the noise power is quite homogeneous in Galactic latitude, this also allows us to evaluate the success of the \focus algorithm from large to small signal over noise ratio.
In practical terms, the masks are taken into account in the averages over sky pixels in Eqs.~\eqref{eq:s1pwst} and~\eqref{eq:s2\cwst_bis}. Thus, the \focus algorithm simultaneously optimizes 
the three Loss terms for each of the five masks, i.e. in total 15 constraints.

Numerically, the optimization runs for 500 iterations between each computation of the noise-induced biases. The minimisation does not improve much and the change in $\tilde{s}$ are negligible after this number of iterations. This step is repeated 12 times (6000 iterations in total), at which point the modification of the estimated biases are very small. The total iteration time represents 10 hours on 3 nodes (processor=Intel Xeon E5-2680) with 28 CPU cores, or 2 hours on 3 M100 GPUs. We also note that the optimization was done on $d-u$ rather than $u$, which was more stable and leads to much less oscillations between local minima. We believe that it is due to the fact that the $d-u$ contamination is close to a Gaussian random field as scales where \focus works, whose pixels values can be optimized in a much more independent way.

\begin{figure*}[tb]
\centering
\includegraphics[width=\textwidth]{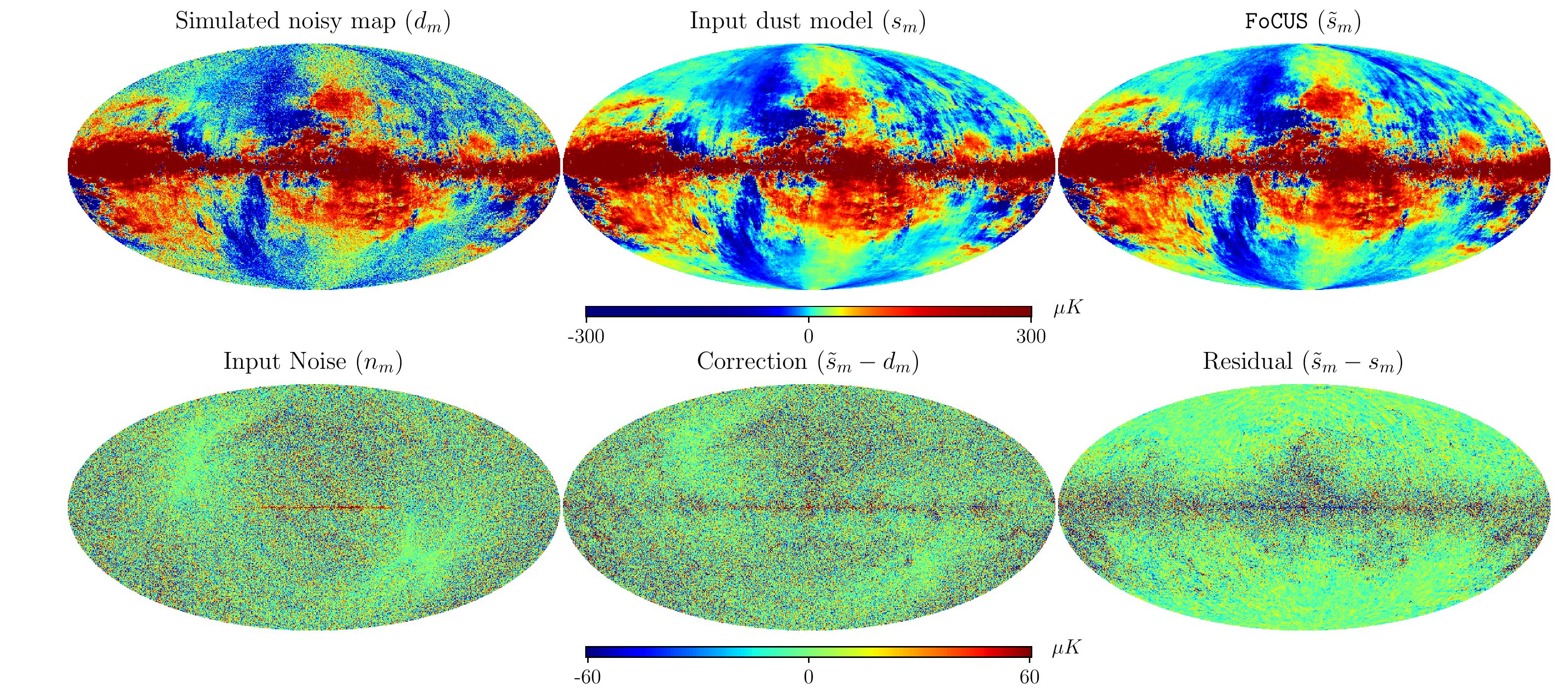}
\caption{Stokes $Q$ maps illustrating the validation on mock data. From left to right on the top row: one relaization of the noisy mock data, the noise-free Vansyngel model and the result of the \focus algorithm. On the bottom line, from left to right the noise map used in the data simulation, the correction found by the \focus method and the residual map: the difference between the Vansyngel model and the \focus map.}
\label{fig:SIM_MAP}
\end{figure*}

\section{Validation of the components separation}
\label{sec:test_simu}

In this section, we apply the \focus components separation method to mock data to assess its performance. We introduce the mock data in Sect.~\ref{subsec:mock_data}, and present the results of the \focus run in Sect.~\ref{subsec:valwst}.
In Sect.~\ref{subsec:valloss}, we analyze the impact of each of the three terms of the Loss function on the \focus output maps. 

\subsection{Mock data}
\label{subsec:mock_data}

To build our mock data set, we combine a model of dust polarization maps with noise simulations. 
We use the dust model, hereafter the Vansyngel model, introduced in Appendix~\ref{sec:AppendixA} of~\citet{Planck2018_III}. The Vansyngel model was used by both~\citet{Planck2018_III} and~\citet{delouis2019} to build end-to-end simulations of the \Planck\ polarization data at 353\,GHz. 
The total intensity maps is the \Planck\ map at 353\,GHz obtained by applying the Generalized Needlet Internal Linear Combination (GNILC) method of~\citet{Remazeilles11} to the 2015 release of \Planck\ HFI maps (PR2). 
The Stokes $Q$ and $U$ maps are built from one realization of the statistical model of~\citet{Vans2017}. In the model of~\citet{Planck2018_III}, the simulation was replaced by the Planck PR2 353\,GHz maps near the Galactic plane, and the largest angular scales $\ell < 20$ of the simulated maps were also replaced by the \Planck\ data (see~\citet{Planck2018_III} for more details).

Away from the Galactic plane and for multipoles $\ell > 20$, the statistical properties of the $Q$ and $U$ maps are those of the~\citet{Vans2017} model. This model is built from a simplified description of the magnetized interstellar medium where the random component of the magnetic field is represented by Gaussian fields.
The $TE$ correlation and $E/B$ asymmetry correlation are introduced in the model maps in spherical harmonics with random phases. However, the model $Q$ and $U$ maps do have some non-Gaussian characteristics that arise
from the total intensity map and the modelling of the line of sight integration with a small number of independent emission layers. As explained in Sec.~\ref{sec:focus}, the temperature map used in the third loss term is the SRoll3 map at 857\,GHz, even if the $T$ map used to evaluate the $TE/TB$ correlations is the one from the Vansyngel model at 353\,GHz. This allows us to remain consistent both for testing the \focus algorithm and validating the result obtained.

We associate the dust model $s_m$ (here the $m$ index is related to modelisation and simulation) with noise maps from the end-to-end \srolltwo dataset~\citep{delouis2019}. Ten noise realizations are added to the dust model to generate ten mock $d_m$ maps, and 300 additional ones, all independent, are used for the \focus optimization.
We apply the \focus method on the ten $d_m$ and obtain ten denoised maps $\tilde{s}_m$. 

\subsection{Validation of \focus on mock data}
\label{subsec:valwst}

We assess the \focus algorithm comparing the maps $\tilde{s}_m$ with the input dust maps in Sect.~\ref{subsubsec:maps}, and their \cswst 
statistics in Sect.~\ref{subsubsec:stats}.

\subsubsection{\focus maps}
\label{subsubsec:maps}

The top row of Fig.~\ref{fig:SIM_MAP} presents three $Q$ maps, from left to right the noisy mock data ($d_m$), the input dust model ($s_m$) and the \focus output $\tilde{s}_m$. The bottom row shows the noise map included in $n_m$, the noise estimate from \focus $d_m-\tilde{s}_m$ and the residual $\tilde{s}_m-s_m$.
The \focus map $\tilde{s}_m$ is strikingly less noisy than $d_m$. It is also clear from Fig.~\ref{fig:SIM_MAP} that the noise estimate $d_m-\tilde{s}_m$ corresponds to what we expect: a noisy map with large scale patterns close to the one of the true noise. To the eye, the residual map appears noisier where the dust emission is the brightest. Along the Galactic plane, the residuals are larger but they represent a small fraction of the total dust signal. Away from the Galactic plane, one can see that the residuals do not correspond to leftover noise, but mainly result from small displacements of some structures between $\tilde{s}_m$ and the true map $s_m$. 

\begin{figure}[!ht]
\centering
\includegraphics[width=0.5\textwidth]{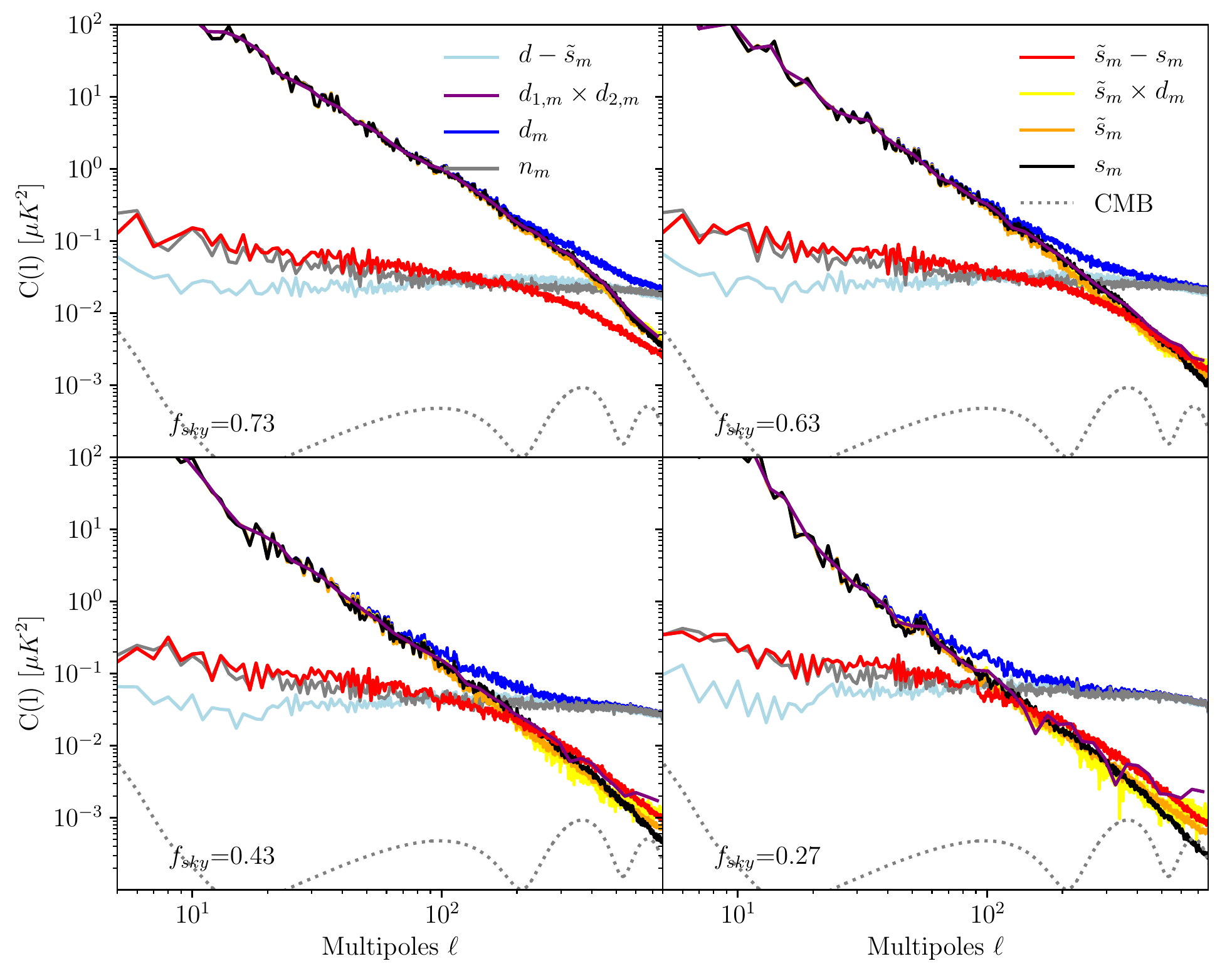}
\caption{Power spectra of one realization for the \focus validation on mock data. The plots show $EE$ spectra of the dust model ($s_m$ in black), the noisy mock data ($d_m$ in blue) and the \focus output ($\tilde{s}_m$ in orange) for four masks with $f_\mathrm{sky}$ from 0.27 to 0.73. The red curve is the power spectrum of the residual map $\tilde{s}_m-s_m$. The figure also presents the noise spectrum ($n_m$) and the noise estimate from \focus ($d_m-\tilde{s}_m$), as well as cross spectra between half-mission mock maps ($d_{1,m}\times d_{2,m}$ logarithmic binned) and between $\tilde{s}_m$ and $d_m$. Note that other mock data realisation show very consistent power spectra.}
\label{fig:SPEC_SIM_MAP}
\end{figure}

In Fig.~\ref{fig:SPEC_SIM_MAP}, we present $EE$ power spectra of the maps in Fig.~\ref{fig:SIM_MAP} for four masks with $f_\mathrm{sky}$ from 0.27 to 0.73. The corresponding $BB$ spectra are shown in Fig~\ref{fig:SPEC_SIM_MAP_B} in \ref{sec:AppendixA}. In both figures, it is remarkable that the power spectra of the \focus output $\tilde{s}_m$ in orange closely follows the true dust spectrum in black in the four plots, down to two orders of magnitude below the noise power for $f_\mathrm{sky} = 0.27$. The success in reproducing deterministic dust structures is characterised by the spectra of the residual map $\tilde{s}_m-s_m$ in red. The power of the residual map becomes larger than that of the dust model in black where the $EE$ power of dust is one tenth of that of the noise. For lower signal-to-noise ratios, the amplitude of the residual spectrum is about twice that of the dust power, which indicates that structures in the \focus output map $\tilde{s}_m$ are not spatially coincident with those in the input map $s_m$. We show below that this corresponds to a regime where the recovered structures have the correct statistical properties, but do not reproduce the input data from a deterministic point of view. The presence of such a regime has already been identified in~\citet{blancard2020statistical}.

\subsubsection{\cswst statistics}
\label{subsubsec:stats}

A main objective of \focus is to derive from an observation a statistical model of dust polarization unbiased by data noise. The mock data allow us to assess the success of the algorithm in this regard. 

\begin{figure}[!ht]
\centering
\includegraphics[width=0.5\textwidth]{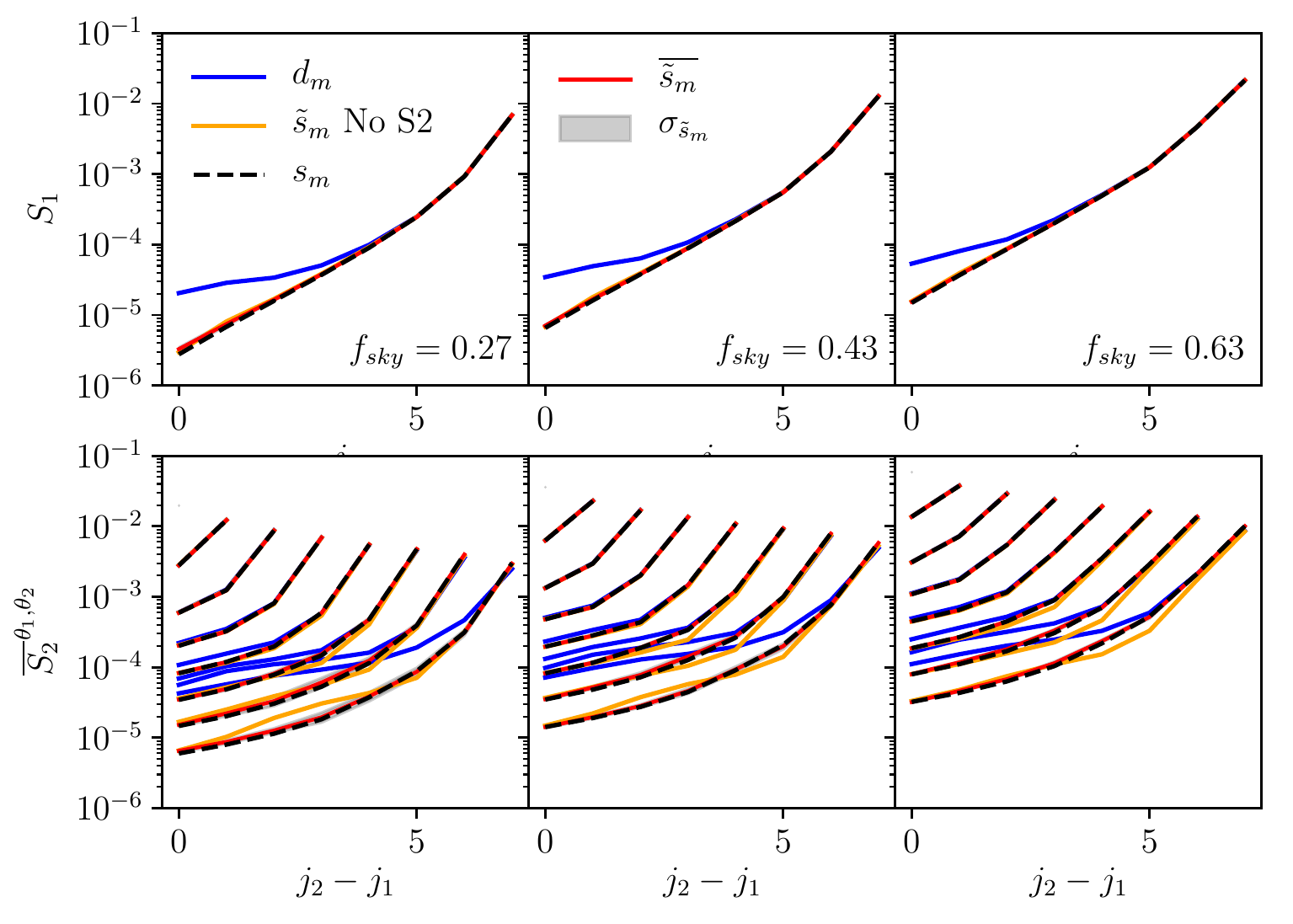}
\caption{Comparison of \cwst\ statistics. The \cwst\ coefficients of the \focus output map $\tilde{s}_m$ (red solid line) are compared to those of the mock data $d$ (blue solid line) and the noise-free dust model $s_m$ (black dashed line). The thicker grey line 
represents the statistical variance of the \focus output estimated from 10 simulations. The top row shows the coefficients $S_{1}$ plotted versus scale $j_1$, and the bottom row the $S_{2}$ coefficients averaged over $\theta_{1}$ and $\theta_{2}$ plotted versus the ratio between the two scales $j_2 - j_1$. The three columns correspond to different Galactic masks.}
\label{fig:SIM_SPEC}
\end{figure}

In Figure~\ref{fig:SIM_SPEC}, we compare the \cwst\ coefficients of the \focus output map $\tilde{s}_m$ to those of the mock data $d_m$ and the noise-free dust model $s_m$. As discussed in Sec.~\ref{sec:valcswst}, the \cwst\ coefficients of a single map boil down to the standard WST coefficients studied for instance in~\cite{Allys_2019}. The top row shows the coefficients $S_{1}$ averaged over $\theta_{1}$ plotted versus scale $j_1$, and the bottom row the mean $S_{2}$ coefficients averaged over $\theta_{1}$ and $\theta_{2}$ plotted versus the ratio between the two scales $j_2 - j_1$, for $j_1 =0$ to 6 from the bottom-right to the top-left. 

The statistics of $d_m$ clearly depart from those of $s_m$. As expected, the difference is most noticeable at small scales, as well as for the area of the sky with the lowest signal-to-noise ratio at high Galactic latitudes. For $f_\mathrm{sky} = 0.27$ and $j_1 = 0$, the mean $S_{2}$ coefficients for $d$ depart from those of $s$ for all $j_2$ but the largest scale. Extracting non-Gaussian statistics of $s_m$ from the noisy $d_m$ data down to scale $j_1=0$ is therefore a notable challenge at high Galactic latitudes, where power ratio between the dust signal and noise ratio is down to 1\% (see Fig.~\ref{fig:SPEC_SIM_MAP}).
In this challenging context, the excellent match between coefficients for $\tilde{s}_m$ and $s_m$ for all masks demonstrates the remarkable success of the \focus algorithm in synthesising maps with the same statistics as the noise-free dust emission. 

For the $353\,$GHz \Planck\ data, we however expect some non-negligible bias to appear on the smallest scales probed with $N_\text{side} > 256$, entering a regime where even the statistical properties of the denoised map begin to differ to those of the true map. This third regime, which has also been observed in~\cite{blancard2020statistical}, is indeed expected in the limit where the noise has a much higher level than the signal. It could however be possible to use an extrapolated model of the \cwst\ statistics of the dust to describe scales included in this regime.

\subsection{Impact of each loss term}
\label{subsec:valloss}

Our loss functions comprise three terms Loss$_{1}$, Loss$_{2}$ and Loss$_{3}$ defined in Eqs.~\ref{eq:loss1}, \ref{eq:loss2} and \ref{eq:loss3}.
The mock data allow us to assess the impact of pairwise combinations excluding one of the loss terms for all sky masks on the \focus output maps. 

\begin{figure}[!ht]
\centering
\includegraphics[width=0.5\textwidth]{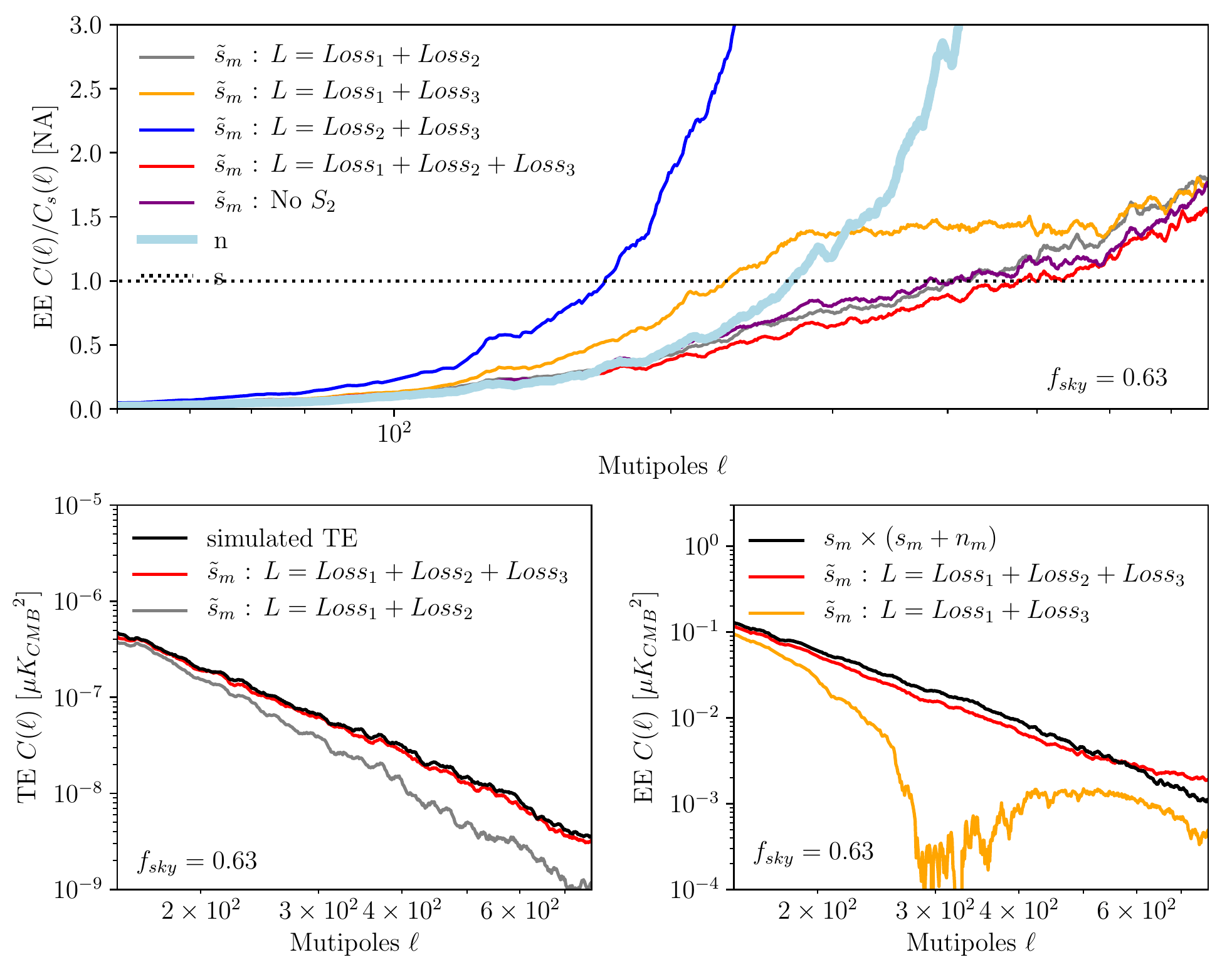}
\caption{Power spectra of the \focus maps for different combinations of the three loss terms Loss$_1$, Loss$_2$ and Loss$_3$. 
The top plot shows the ratio between the $EE$ spectra of the residual map, $s-\tilde{s}_m$, and the input dust model $s_m$. The bottom left plot compares the $TE$ spectra obtained with or without Loss$_{3}$, in red and grey colours respectively. 
The bottom right plot compares the cross spectra between \focus map $\tilde{s}_m$ and the mock data $d$ with or without 
Loss$_2$, in red and yellow colours respectively. All of the spectra are binned in $\ell$ bins with a width $\Delta \ell = 10$.}
\label{fig:L1L2}
\end{figure}

Figure~\ref{fig:L1L2} include three plots where the input dust model $s_m$ is drawn in black.
All plots are for $f_\mathrm{sky} = 0.63$.
The top plot presents the $EE$ spectra of the residual maps $\tilde{s}_m-s_m$ divided by that of $s_m$.
This ratio quantifies the ability of the \focus algorithm to reconstruct structures consistently with the noise-free dust model.
The bottom left plot compares the $TE$ spectrum obtained with or without 
the Loss$_3$ term, in red and grey colours. 
The bottom right plot compares the cross spectra between
the FoCUS map $\tilde{s}_m$ and the mock data $d_m$ with or
without the Loss$_2$ term in red and yellow colours, respectively.

In the three plots, the \focus output maps $\tilde{s}_m$ obtained with the complete loss function, drawn in red colour, 
is the best result, which confirms our choice of combining the three terms of the loss function. 
The top plot shows that both Loss$_1$ and Loss$_2$ are essential to minimize the power of the residual maps\footnote{We interpret the effect of Loss$_2$ as follow. At multipoles between $10^2$ and $3\cdot 10^2$, this loss allows to recover the deterministic structures which can be characterized from the cross-statistics with the noisy-data. At higher multipoles, these cross-statistics progressively become noise-dominated, and the lever-arm of this loss term to recover deterministic structures decreases.}.
Loss$_2$ also ensures a faster convergence of the minimization. Furthermore, 
the bottom right plot shows that Loss$_2$ is critical to match the cross spectra $\tilde{s}_m\times d_m$. 
Loss$_3$ has only a weak impact on the residual maps, but the bottom left plot shows
that it is essential to match the $TE$ correlation of the dust model, as expected because it constrains the cross statistics 
between dust polarization and total intensity. 

The top plot in Fig.~\ref{fig:L1L2} also shows with purple colour the residual power for $\tilde{s}_m$ maps obtained discarding the $S_{2}$ coefficients in the loss function. Comparing the purple and red curves, it is interesting to see that those coefficients only improve by a few tens of percents the residual power. However, yellow curve in Fig.~\ref{fig:SIM_SPEC} demonstrates that the $S_{2}$ coefficients in the loss terms are important to fully recover the non-Gaussian properties of the dust map, as expected.

\begin{figure*}[!ht]
\centering
\includegraphics[width=\textwidth]{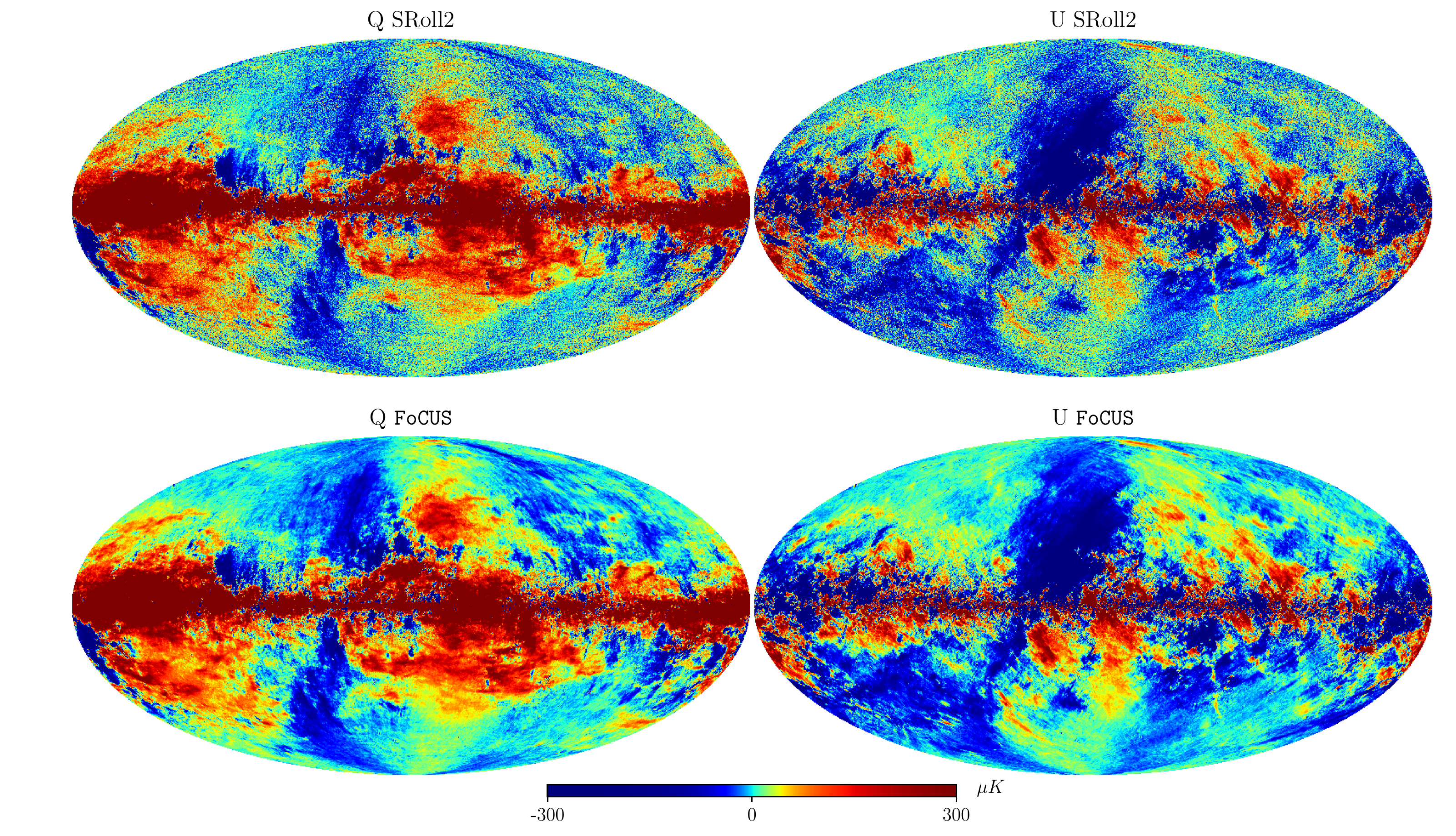}
\caption{Stokes $Q$ and $U$ maps at 353\,GHz for $N_\mathrm{side} = 256$. The top images present the input \Planck\ \srolltwo maps and the bottom ones the corresponding \focus maps.} 
\label{fig:LOCAL_MAP}
\end{figure*}

\section{Denoised Planck dust polarization maps}
\label{sec:result_data}

In this section, we use the CWST and \focus to separate dust polarization and data noise in \Planck\ data. The data we use is introduced in Sect.~\ref{subsec:input-data}, and the denoised \Planck\ polarization maps are presented in Sect.~\ref{subsec:focus_maps}.

\subsection{Input data}
\label{subsec:input-data}

We use the \Planck\ Stokes $Q$ and $U$ maps at 353\,GHz from the \srolltwo\ processing, 
which corrects instrumental systematics present in the \Planck\ Legacy polarization maps of the High Frequency Instrument\footnote{The \srolltwo\ maps are available \href{http://sroll20.ias.u-psud.fr/sroll22_files.html}{here.}}~\citep{delouis2019}. To obtain dust polarization maps, we subtract the CMB polarization using $Q$ and $U$ maps from the SMICA components separation method~\citep{compsep2018}. Over the multipole range we consider, at 353\, GHz dust polarization dominates the CMB signal for all sky masks as shown in Fig.~\ref{fig:SPEC_SIM_MAP}. Uncertainties on the CMB correction are thus not an issue. As explained in Sec.~\ref{sec:focus}, the temperature map used in the third loss is the SRoll3 map at 857\,GHz.

\subsection{The \focus maps}
\label{subsec:focus_maps}

\begin{figure*}[!ht]
\centering
\includegraphics[width=\textwidth]{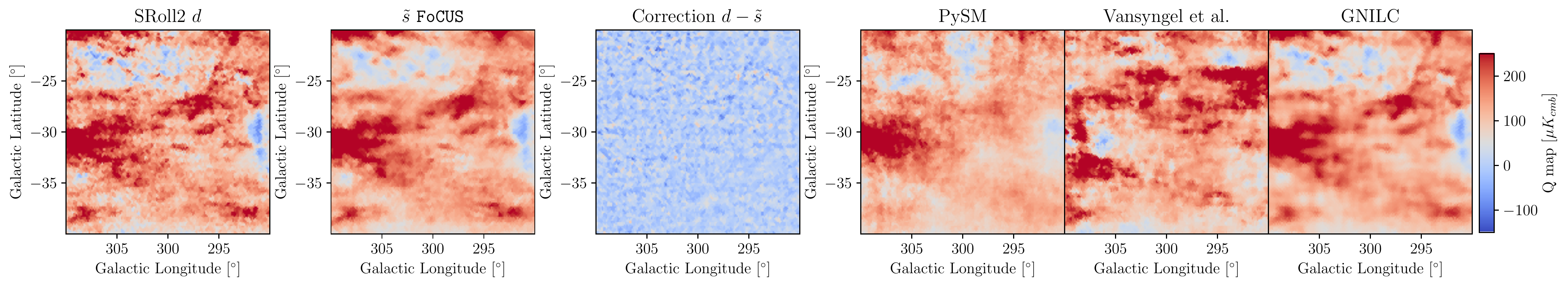}
\caption{Zoom on a sky region of the Q map. From the left to right images are the SRoll 2.0 data, the \focus map, The correction computed by \focus to be applied to the SRoll2.0 map, the PySM d1 model, the~\citep{Vans2017} map and the GNILC map.}
\label{fig:COMP_MAP}
\end{figure*}

Figure~\ref{fig:LOCAL_MAP} presents the \srolltwo\ Stokes $Q$ and $U$ maps at  353\,GHz (top images) and the result of the \focus algorithm (bottom images). The eye-comparison shows that noise has been subtracted without smoothing the map. This is further illustrated in Fig.~\ref{fig:COMP_MAP}, which zooms on one sky area to allow a detailed comparison with other dust polarization models\footnote{Note that even without an over-smoothing, the dust map has much less power at small scales than the noise. It is then expected for the denoised map to have a smoother texture.}. 

The PySM d1 model map~\citep{Thorne17,Zonca21} includes small scale structure, derived from a random Gaussian field, which appears unrealistic. The Vansyngel maps are constructed from the dust total intensity map. Their textures seem closer to what is statistically expected but the small scales seem to lack of elongated structures. GNILC map shows lack of small scales. Finally, the \focus map shows non-Gaussian elongated structures at all scales, a general characteristics of the diffuse interstellar medium, which our algorithm is able to capture thanks to the use of advanced statistical descriptors and constraints.

\begin{figure}[!ht]
\centering
\includegraphics[width=0.5\textwidth]{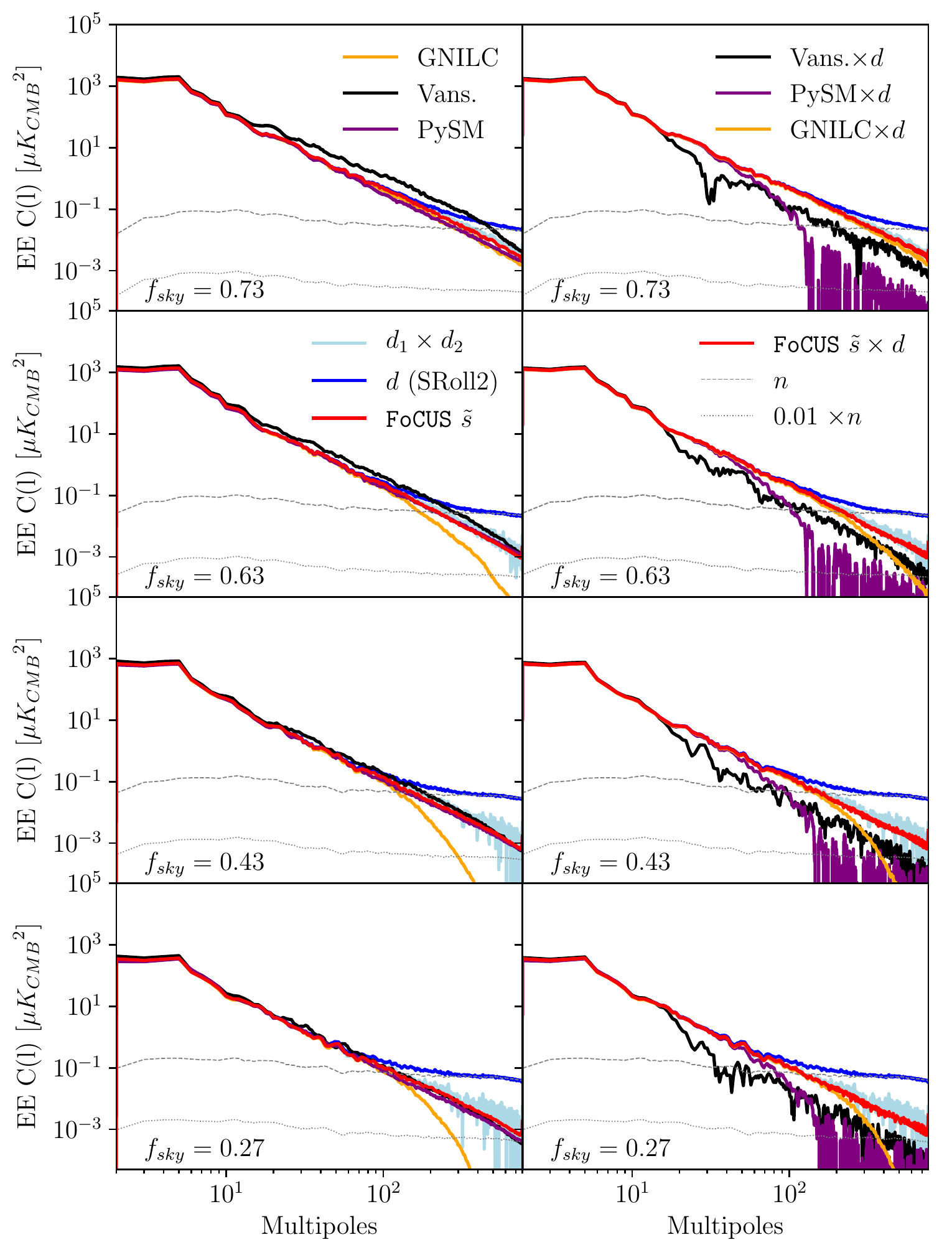}
\caption{$EE$ power spectra (left column) and cross-power spectra (right column) for Galactic masks with $f_{\rm sky}$=0.27 (bottom) to 0.73 (top). In each plot, the cross spectrum of the two \Planck\ half-mission maps is drawn in purple color. 
The blue, orange and red curves represent the power spectra of the SRoll2, GNILC and \focus maps, respectively.
All the spectra are binned over 10 multipoles and normalized by dividing the power with the $f_{\rm sky}$ value to keep the scales consistent between the plots.}
\label{fig:EE_SPEC}
\end{figure}

Power spectra of the \Planck\ data are compared with those of the \focus,  GNILC, PySM, and Vansyngel maps are compared, for four Galactic 
masks defined by their corresponding $f_{\rm sky}$, in Fig.~\ref{fig:EE_SPEC} and Fig.~\ref{fig:BB_SPEC} (in appendix) for $EE$ and $BB$, respectively.
In each figure, plots in the left column present power spectra and those in the right column cross-spectra with the \srolltwo\ input maps. Each plot includes 
the cross spectrum between the two \srolltwo\ half-mission maps ($d_1\times d_2$) drawn in purple color, as the reference to match because it is a noise-unbiased estimate from the \Planck\ data of the spectrum of Galactic dust emission. 

In the left column of the two figures, the \srolltwo\ spectra show the data noise bias over an increasing range of $\ell $ as $f_{\rm sky}$ decreases. It is remarkable that the power spectra of the \focus map is consistent with the reference for the four masks, up to multipoles where the signal power is two orders of magnitude lower than that of the noise. 
The difference with the GNILC method, which reduces the sky noise at the expense of small scales smoothing, stands out in Fig.~\ref{fig:EE_SPEC}. 
The plots also show that the power spectra of the Vansyngel maps deviate somewhat from the \Planck\ data, which is a known shortfall of their model. 

Plots in the right columns of Fig.~\ref{fig:EE_SPEC} and Fig.~\ref{fig:BB_SPEC} present cross-power spectra between models and the \srolltwo maps. Our purpose is to quantify the correlation on the sky between the data and model maps, but we note that our validation of the \focus method has shown (see section~\ref{sec:valcswst}) that some correlation may arise from data noise. On these plots, like for those in the left columns, the $d_1 \times d_2$ cross-power spectra are the references to match. Thus, it is satisfactory to see that the cross-power spectra between the \focus and \srolltwo maps is close to $d_1 \times d_2$ for the four masks. The match is excellent for $f_{\rm sky} = 0.73$. For the other masks we see some loss of correlation at high $\ell$, which increases for decreasing $f_{\rm sky}$, i.e. for decreasing signal-to-noise ratio.
The correlation is for all masks larger than that measured for the PySM and Vansyngel models. This is expected because both these models are constructed with an algorithm that is not designed to preserve correlation with the input data. 

\subsection{TE and TB correlation }

\begin{figure}[!ht]
\centering
\includegraphics[width=0.5\textwidth]{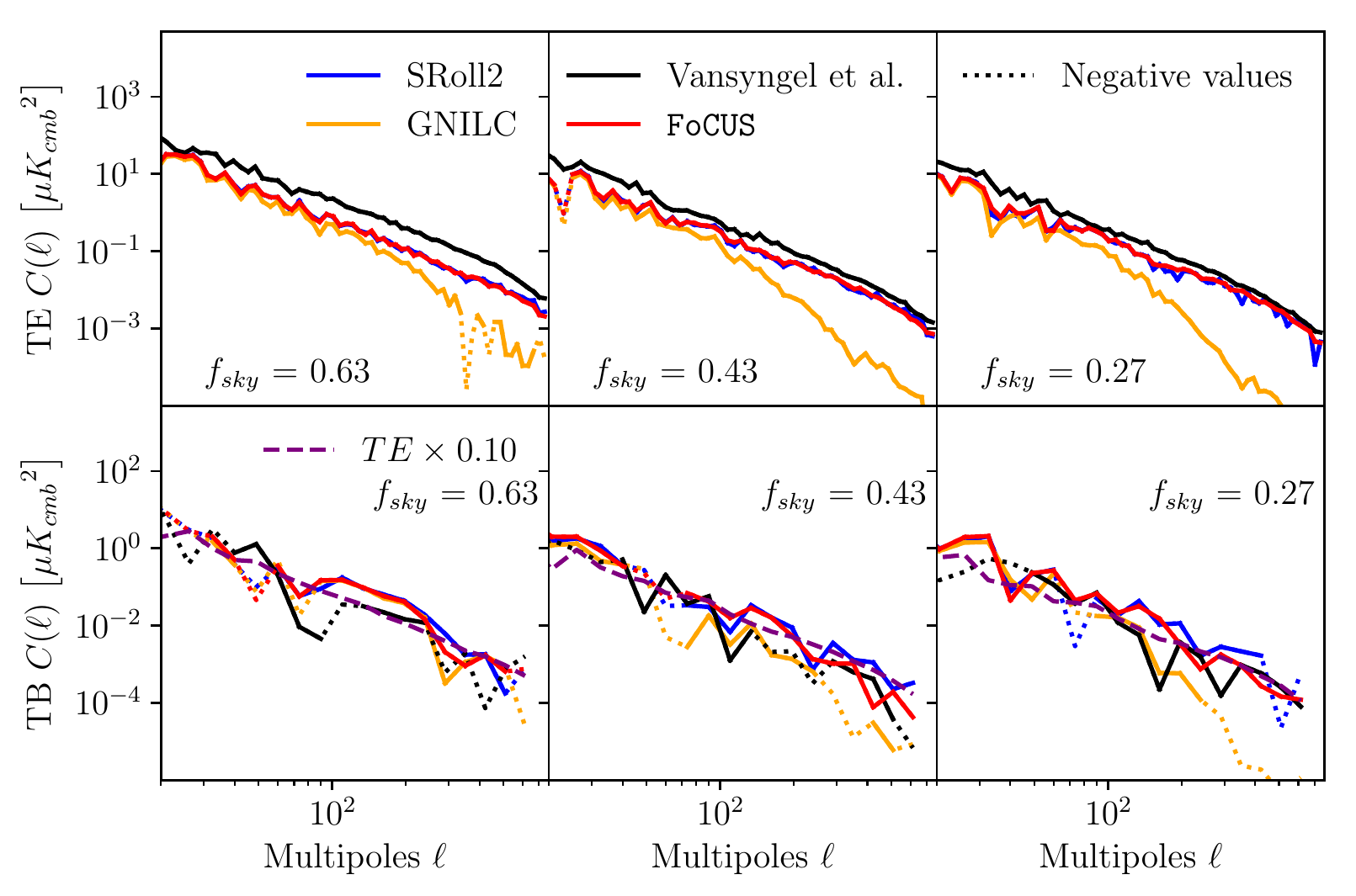}
\caption{$TE$ (top row) and $TB$ (bottom row) cross-power spectra of the \srolltwo  (blue curves) and \focus maps (red curves). The orange and black curves show the same results for the 
GNILC and Vansyngel maps. Each column corresponds to a different galactic mask. The cross-power spectra are binned in bins of width $\frac{\Delta \ell}{\ell} = 0.05$ for $TE$ and $0.2$ for $TB$ to reduce the noise variance. Doted lines represent negative values.}
\label{fig:TETB}
\end{figure}

The $TE$ and $TB$ correlation are major statistical characteristics of dust polarized emission, determined by the analysis of \Planck\ data~\citep{Planck2018XI}. This important property is difficult to match without learning the model statistics from data. 

In Fig.~\ref{fig:TETB}, the $TE$ and $TB$ cross-power spectra for the \srolltwo and \focus maps are compared. The plots show that the two sets of $TE$ and $TB$ cross-power spectra match, and that the \focus\ algorithm reduces the variance at the highest $\ell$. This is a clear success of the \focus\ algorithm. The Fig.~\ref{fig:TETB} also shows that the previous noise removal (see GNILC or~\citep{Vans2017} in orange or black) do not reproduce these correlations.

\section{Projected applications and perspectives}
\label{sec:prospects}

The \Planck\ 353\,GHz polarization maps have been extensively used for 
the astrophysics of dust polarization and the modelling of the Galactic dust foreground to CMB.
For both topics, the \focus map represents a significant stepping-stone opening new prospects.
We illustrate and discuss these perspectives from both the astrophysics and foregrounds viewpoints.

\subsection{Astrophysics of dust polarization}
\label{subsec:astrophysics}

For astrophysics, the signal-to-noise ratio of the dust polarization maps statistically conditions the range of angular scales accessible to study. The polarized emission of the diffuse ISM at high Galactic latitudes is too faint to be analyzed at the full $5'$ angular resolution of \Planck (see dust power spectra in \citet{Planck2018XI}). As a matter of fact, most of the analysis of the dust polarized emission in~\citet{Planck2018XII} was performed after smoothing the \Planck\ maps to 80' and even 160' resolution. 
The \focus maps thus allow us to extend the range of earlier studies. As our data denoising is statistical in nature at scales where signal-to-noise ratio is low, the \focus maps are most relevant to statistical studies, in particular those that aim at characterizing statistically the turbulent component of the Galactic magnetic field.

\begin{figure*}[!ht]
\centering
\includegraphics[width=\textwidth]{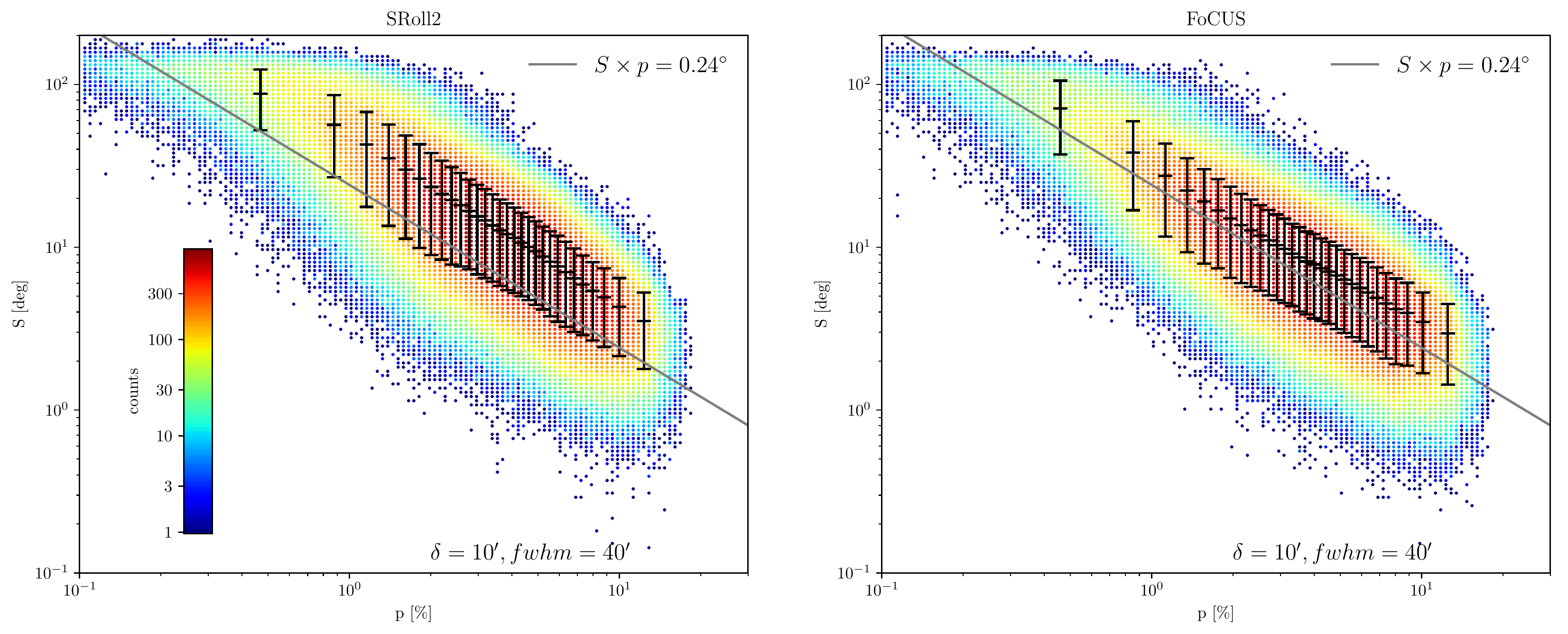}
\caption{Distribution of the polarization angle dispersion $\mathcal{S}$ and the polarization fraction $p$ values for the \srolltwo (left panel) and \focus (right panel) maps. The data points with error-bars represent the mean value and $1\sigma$ dispersion of $\mathcal{S}$ within bins of $p$. The data binning and representation follows that of Fig.~10 of~\citet{Planck2018XII}. 
The gray line shows the relation $\mathcal{S} \times p = 0.24^\circ$, which is expected based on the analytical model of~\citet{Planck2018XII} for the resolution FWHM=40' we use.}
\label{fig:FIG10}
\end{figure*}

To illustrate this perspective, we use the \focus\ maps to compare the polarization angle dispersion, $\mathcal{S}$, and the polarization fraction, $p$, as done by~\cite{Planck2018XII} with the \Planck\ Legacy maps. $\mathcal{S}$, introduced by~\cite{Hildebrand09} and~\cite{planck2014-XIX}, quantifies the local
non-uniformity of the polarization angle patterns on the sky by means of the local variance of the polarization angle map on a scale defined by a lag $\delta$. Regions where the polarization angle tends to be uniform exhibit low values of $\mathcal{S}$, while regions where the polarization patterns change on the lag-scale exhibit larger values. The polarization fraction, $p$, depends on both the orientation of the mean Galactic magnetic field in the Solar Neighborhood and, statistically, on depolarization resulting from changes in the magnetic field orientation along the line of sight~\citep{Planck_PIP44}. 
The trend $\mathcal{S} \propto 1/p$ observed in the \Planck\ data can be reproduced with an analytical model presented in Appendix~\ref{sec:AppendixA} of~\citet{Planck2018XII}. The product $\mathcal{S} \times p$ scales with the degree of randomness of the Galactic magnetic field: the ratio between the dispersion of the turbulent component and the strength of the mean field. 

Figure~\ref{fig:FIG10} presents plots of $\mathcal{S}$ versus $p$, for both the \srolltwo\ and \focus\ maps, with the same presentation as in the corresponding Figure (also Fig.~10) in~\citet{Planck2018XII}. Figure~\ref{fig:FIG10sim} presents the corresponding plots for the Vansyngel model.
Our maps have a resolution of 40' (FWHM) to be compared with 160' for the plot in~\citet{Planck2018XII}. 
As in~\citet{planck2014-XIX}, we use a lag equal to one fourth of FWHM, i.e. $\delta = 10'$. 
The model relation for our resolution and lag, $\mathcal{S} \times p = 0.24^\circ$, is the line drawn in both plots. We see from Fig.~\ref{fig:FIG10sim} that the noise bias on the $\mathcal{S}$ and $p$ relation is considerably reduced by the \focus algorithm. Given the excellent debiasing obtained on this validation dataset, the \focus\ plot in Fig.~\ref{fig:FIG10} suggests that the analytical model of~\citep{planck2014-XIX} only provides an approximate description of the data.

\subsection{Galactic foregrounds modelling}
\label{subsec:foregrounds}

The modelling of the dust foreground to CMB polarization is the primary motivation of our work. 
The approach we follow to model dust polarization is novel in some crucial aspects, which we summarize here.

\begin{figure}[!ht]
\centering
\includegraphics[width=0.5\textwidth]{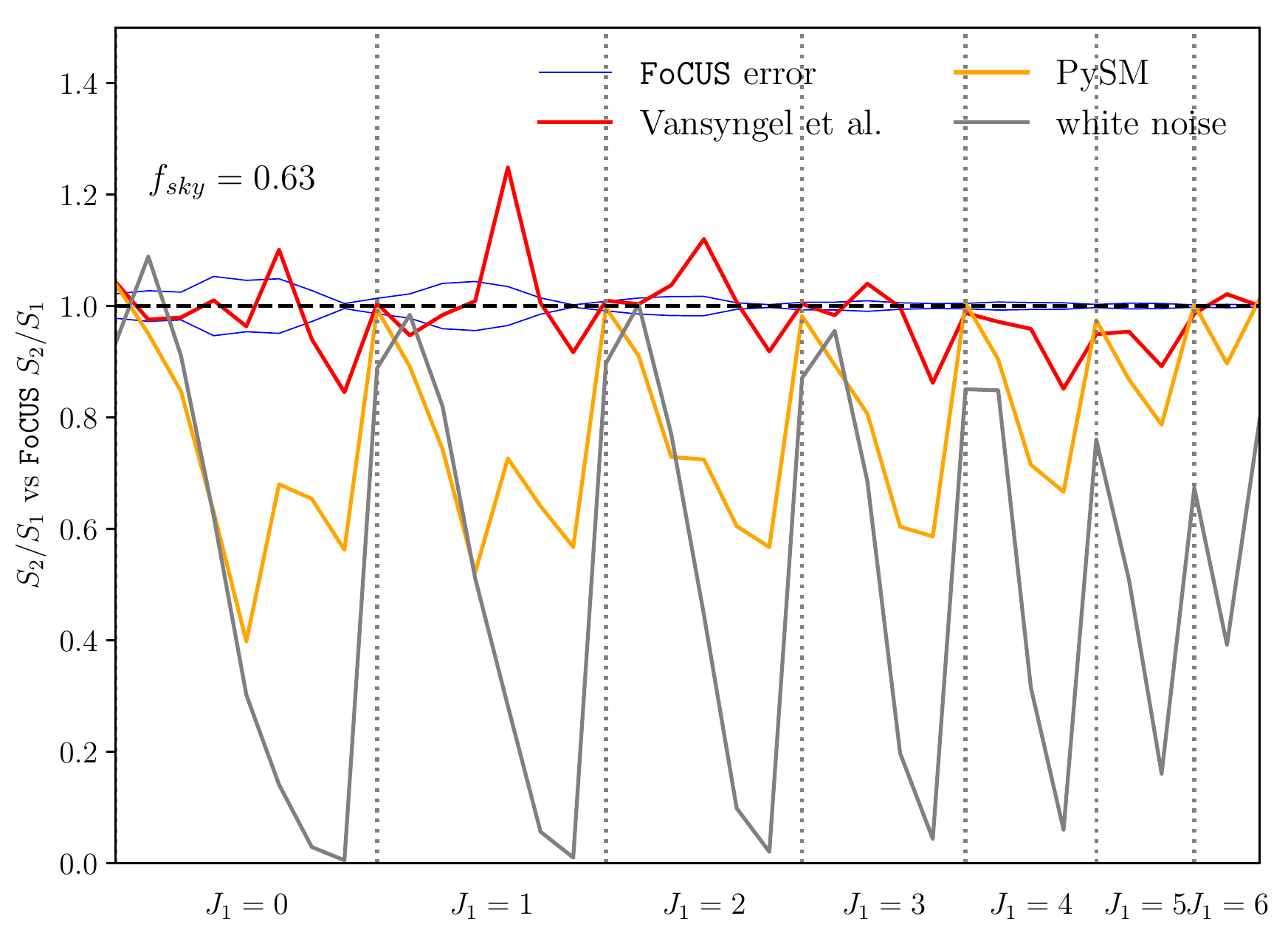}
\caption{Comparison of the second order $S_2$ \cwst coefficients used as a measure of non-Gaussianity. The $S_2$ coefficients of $E$ maps normalized by $S_1$ are plotted versus scales $J_1$ and $J_2$. The data are ordered by $J1$ values and for each $J1$ for increasing $J2$. The plot compares $S_2/S_1$ ratios for a Gaussian map in purple, PySM model in yellow and that of~\citet{Vans2017} and the \focus map with a black dashed line. All data values are normalized by those of the \focus map. The two blue curves shows the uncertainty ($\pm 1\sigma$) on the $S_2/S_1$ ratio estimated from simulations. }
\label{fig:COEF_COMP}
\end{figure}

We learn our statistical model from the \Planck\ data using a set of summary statistics designed to perform an in-depth characterization of non-Gaussian structures~\citep{bruna2013}. The \focus maps include non-Gaussian features that are missed by models making use of Gaussian random fields to describe foregrounds on scales where the \Planck\ template maps are noise dominated. This important difference is illustrated in Fig.~\ref{fig:COEF_COMP}, where ratios between \cwst\ coefficients $S_2/S_1$, used as a measure of non-Gaussianity, 
are compared between the \focus\, PySM and Vansyngel $E$ maps.

We have applied to all-sky \Planck\ maps a statistical components separation that allow us to extend our statistical model of dust polarization down to scales where the dust power is two orders of magnitude smaller than the data noise. For the sky at high Galactic latitudes best suited for deep CMB observations we succeed in modelling dust polarization up to $N_\mathrm{side}=256$. The resolution could be increased by modelling and extrapolating the scale-dependence of \cwst\ coefficients.

The \cwst\ statistics allow us to combine different maps. We have used this possibility to learn from the \Planck\ data the correlation between dust total intensity and polarization. Therefore, our \focus map reproduces the $E/B$ asymmetry, the $TE$ and also $TB$ correlations, a unique achievement among current dust models illustrated in Fig.~\ref{fig:TETB}. This figure also shows that denoising methods like GNILC, which seek to minimize the difference between the denoised map and the true signal, remove power on scales dominated by the noise.

To go further on the modelling of polarized Galactic foregrounds, one main objective will be to extend this analysis to multi-frequency models that take into account the spatial variations of the spectral energy distribution of dust polarization~\citep{Ritacco22}. For this purpose, it is necessary to properly construct joint multi-channels Scattering Transform statistics, as well as to extend the components separation algorithm to this framework. 

\section{Conclusion}
\label{sec:conclusion}

We have applied the scattering transform to \planck\ data 
in order to derive a non-Gaussian model of dust polarization and produce 
denoised all-sky dust Stokes $Q$ and $U$ maps at 353\,GHz. 
First, we introduced the \cwst\ statistics that we use to characterize the non-Gaussian structure of dust polarization. They extend the computation of scattering coefficients to the Healpix pixelization on the sphere and include cross-statistics that allow us to combine images. Second, we devised the \focus\ algorithm that uses the \cwst\ statistics to separate dust polarization from data noise. \focus\ is validated on simulations of the \Planck\ data, before being applied to the \srolltwo\ \Planck\ maps at 353\,GHz. The main results of our work are as follows. 

The \cwst\ statistics and the \focus algorithm allow us to characterize dust polarization down to angular scales where the $EE$ dust power is two orders of magnitude smaller than that of the data noise. The \focus Stokes maps reproduce \Planck\ dust polarization power spectra estimated from cross-spectra of half-mission maps over these scales. 

Our validation on mock data allow us to compare the \focus\ output map $\tilde{s}$ with the noise-free input map $s$. The spectra of the residual map  $\tilde{s}-s$ becomes larger than that of $s$ at scales where the $EE$ dust power is lower than one tenth of the noise power. On these scales, structures in the \focus\ output maps $\tilde{s}$ have comparable non-Gaussian statistics, estimated in terms of \cwst,
but are not spatially coincident with those in $s$. 

The \focus\ Stokes maps at 353\,GHz\footnote{The \focus\ maps are available \href{http://sroll20.ias.u-psud.fr/sroll40_353_data.html}{here.}}, with a set of residual maps from our mock data analysis quantifying uncertainties, are made available to the community. The \focus maps open new prospects for astrophysics and the modelling of the Galactic dust foreground to CMB polarization.

For astrophysics, the signal-to-noise ratio of the dust polarization maps limits the range of angular scales accessible to study. Since our denoising of the data is statistical in nature at scales where the signal-to-noise ratio is low, the gain is in statistical studies. We illustrate this type of applications repeating the \Planck\ studies of the anti-correlation between the dispersion of polarization angles and the polarization fraction. 

The \focus\ Stokes maps improve on current models of dust polarization in two main aspects.
(1) The \focus maps include non-Gaussian characteristics of dust polarization, which are missed by models making use of Gaussian random fields to describe foregrounds on scales where the \Planck\ maps are noise dominated.
(2) The \cwst\ cross-statistics allow us to learn from the \Planck\ data the correlation between dust total intensity and polarization. Therefore, our \focus map reproduces the $E/B$ asymmetry and the $TE$ and $TB$ correlations, a unique achievement among current dust models used to assess CMB components separation methods.

A clear path to improve our results would be to use state-of-the-art Scattering Transform statistics on the sphere. Following recent works, several improvements of the scattering statistics could be done in the near future. On the one hand, more refined wavelet transforms on the sphere could be used, as discussed in Sec.~\ref{sec:valcswst}. One the other hand, other successful sets of scattering statistics, which give better syntheses on regular 2D grids, could be used. For instance, the Wavelet Phase Harmonics~\citep{allys2020,Jeffrey22}, and their recent multi-channel extensions~\citep{Regaldo22}, or the more recent representations built from Wavelet Scattering Covariances~\citep{morel2022scale,cheng2022scatcov}. However, the main challenge is to make such improvements feasible given the computational and memory costs of the \focus algorithm.

The scattering coefficients derived from the Planck data could also be used to generate a set of realistic synthetic non-Gaussian foreground maps. Several ways to proceed could be thought of, which would depend on the scientific objective of such a generation. This is an open issue for future works.

On a more general aspect, the \cwst\ statistics and the \focus algorithm could be applied to many processes defined on the sphere, including in other area than astrophysics, to leverage and analyze different types of correlated data. Its two main advantages are its ability to efficiently combine different datasets and statistical constraints.

\begin{acknowledgements}
This work is part of the R$\And$T Deepsee project supported by CNES. The authors acknowledge the heritage of the Planck-HFI consortium regarding data, software, knowledge. This work has been supported by the Programme National de Télédétection Spatiale (PNTS , http://programmes.insu.cnrs.fr/pnts/), grant n° PNTS-2020-08 and by the CNRS to the 80~Prime project AstrOcean. FB acknowledges support from the Agence Nationale de
la Recherche (project BxB: ANR-17-CE31-0022). The authors acknowledge valuable discussions with B. Regaldo on cross-scattering coefficients. The authors also thank C. Auclair, S. Cheng, M. Eickenberg, F. Levrier, J. McEwen, S. Mallat, B. Ménard, R. Morel, and P. Richard for useful discussions.
\end{acknowledgements}

\bibliography{main}

\appendix
\section{Additional figures}
\label{sec:AppendixA}

\subsection{Power spectra for B-modes}

\begin{figure}[!ht]
\centering
\includegraphics[width=0.5\textwidth]{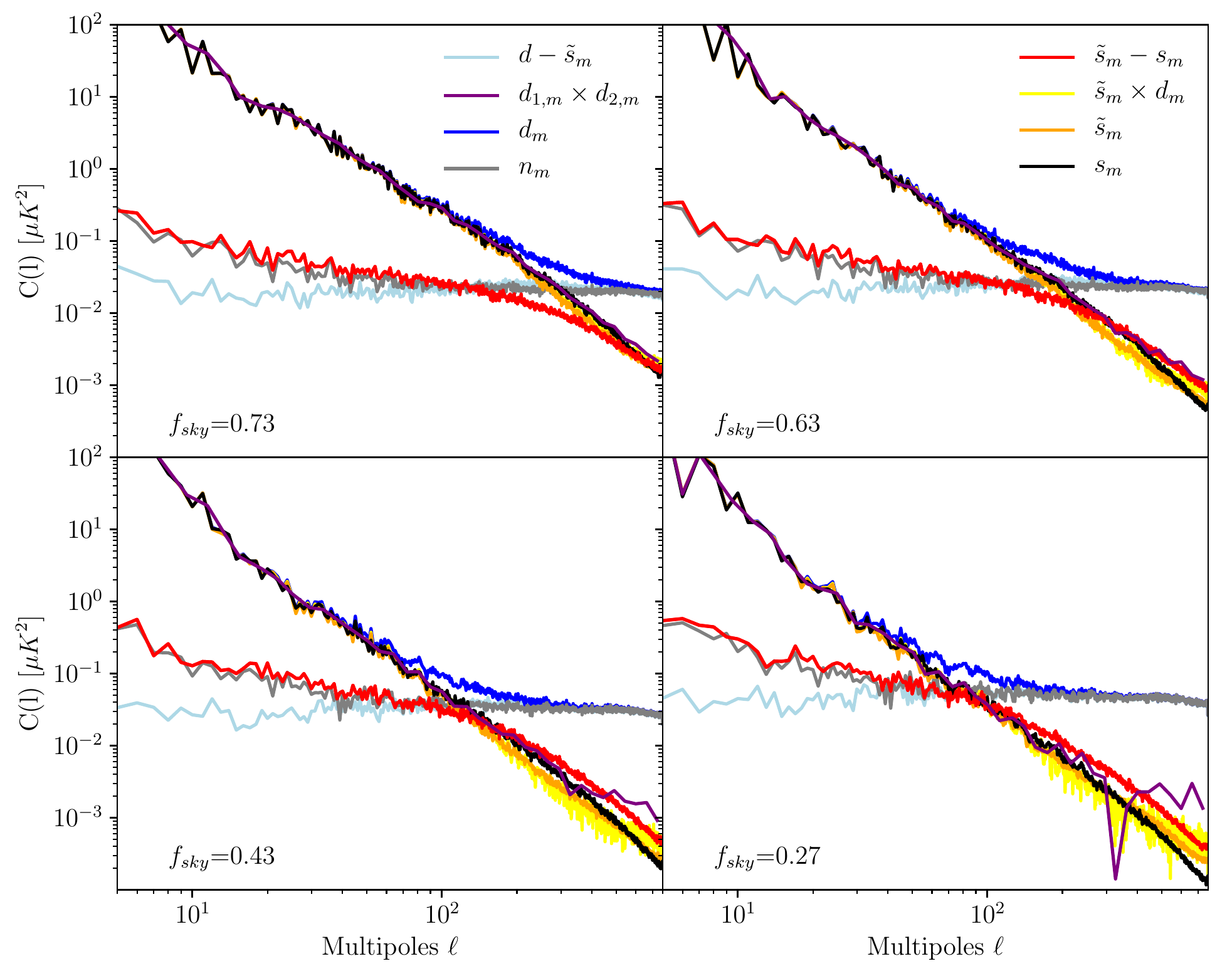}
\caption{Power spectra of one realization for the \focus validation on mock data. The plots are the same as those in  Fig:\ref{fig:SPEC_SIM_MAP} but for $BB$ spectra.}
\label{fig:SPEC_SIM_MAP_B}
\end{figure}

Figure~\ref{fig:SPEC_SIM_MAP_B} complements Fig.~\ref{fig:SPEC_SIM_MAP} by presenting $BB$ power spectra of the \focus validation for one realization of the mock data. The signal-to-noise ratio is lower for $BB$ than $EE$ as the $E$ to $B$ power ratio for dust emission is about 2.  Furthermore, the $TB/TE$ power ratio  is about one tenth, which decreases  the impact of the loss term (Loss$_3$) based on this correlation. For the mock data, Loss$_3$ is fully ineffective because the Vansyngel model does not include the $TB$ correlation. These effects combine to make \focus denoising more challenging. However, Fig.~\ref{fig:SPEC_SIM_MAP} shows a good consistency between the $BB$ power spectra of the \focus\ and the input model maps for most multipoles. At very high Galactic latitudes (bottom right panel), the noisier part of the sky, the $BB$ power spectrum of the \focus maps shows a small bias compared to that of the input maps, which reflects the limitation of our method when the signal-to-noise ratio is very low  ($<1\%$).

\begin{figure}[!ht]
\centering
\includegraphics[width=0.5\textwidth]{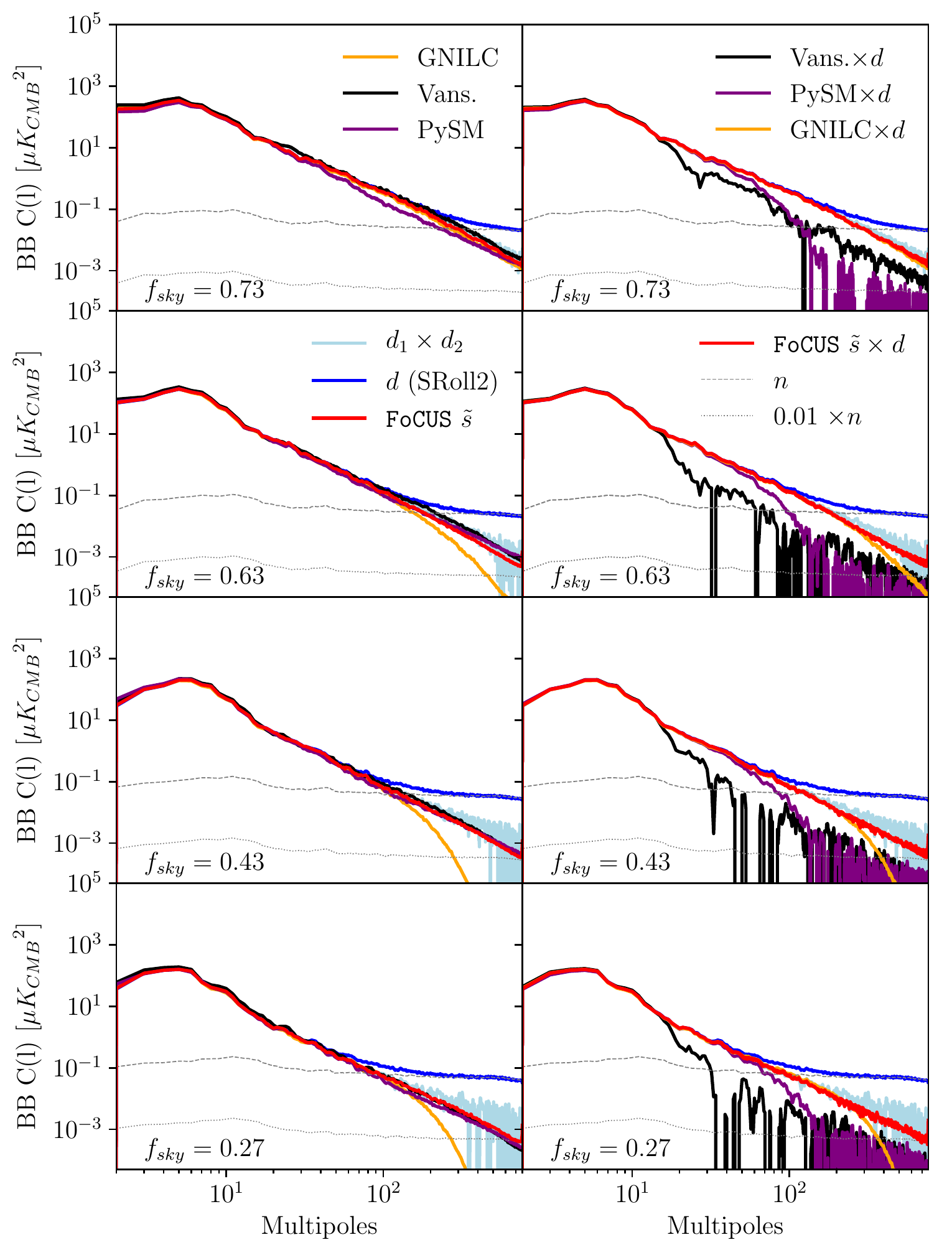}
\caption{Power spectra (left column) and cross-power spectra
(right column) for Galactic masks with $f_\mathrm{sky} =0.27$ (bottom) to
0.73 (top).  The plots are the same as those in   Fig:\,\ref{fig:EE_SPEC} but for $BB$ spectra.}
\label{fig:BB_SPEC}
\end{figure}

Figure~\ref{fig:BB_SPEC} shows the same set of  power spectra as in Fig.~\ref{fig:EE_SPEC} but for $BB$. These plots demonstrate that the \focus results are consistent for $BB$ and $EE$. The fact that both the $EE$ and $BB$ power spectra are properly retrieved demonstrates that the \focus maps  keep the $EE$ to $BB$ power asymmetry.

\subsection{$\mathcal{S}$ and $p$ plot for the mock data}
\begin{figure*}[!ht]
\centering
\includegraphics[width=\textwidth]{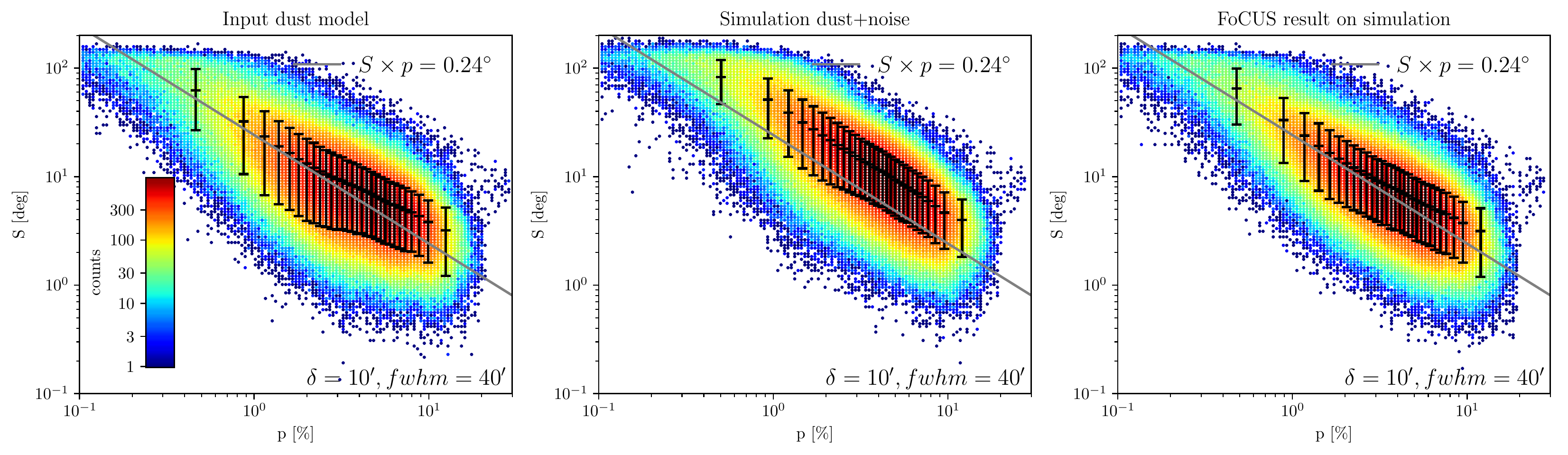}
\caption{Same plot as in Fig.~\ref{fig:FIG10} for the Vansyngel model and the mock data. The figure shows the joint distribution of  $\mathcal{S}$ and $p$  for  the Vansyngel model (left panel), one realization of the noisy mock data (middle panel) and the \focus denoised maps (right panel).}
\label{fig:FIG10sim}
\end{figure*}

Figure~\ref{fig:FIG10sim} complements Fig.~\ref{fig:FIG10} by presenting the joint distribution of  $\mathcal{S}$ and $p$ for  the Vansyngel model (left panel), one realization of the mock data (middle panel) and the result of the \focus denoising applied on the mock data (right panel). The eye-comparison of the three plots show that 
\focus  considerably reduces the noise bias on the  $\mathcal{S}-p$ relation.  We note that the analytical model of \citet{Planck2018XII} does not match the Vansyngel model for high $p$ values. A similar shift between the analytical model and the data is observed for the \Planck\ maps in the right panel of Fig.~\ref{fig:FIG10}.

\end{document}